\documentstyle[12pt]{article}
\renewcommand{\theequation}{\arabic{section}.\arabic{equation}}
\renewcommand{\thefootnote}{\fnsymbol{footnote}}
\newsavebox{\innhamnuber}
\newsavebox{\scalprod}
\newsavebox{\innprod}
\def\pint{\mathbin{\raise1.5pt\hbox{$\underline{\raise-.5pt\hbox
{$\phantom{n}$}}\mskip-2mu\scriptstyle|$}}}
\begin{document}
\language=0
\begin{titlepage}
\begin{flushright}
IC/98/138\hspace*{16.5mm}\hfill\\
hep-th/9809104\hspace*{3.5mm}\hfill\\
10 September, 1998
\end{flushright}
\vspace*{12mm}
\begin{center}
\Large{\bf Hidden supersymmetry and Berezin quantization \\ of N=2, D=3
spinning superparticles}
\vspace{0.8cm}

\large{I.V. Gorbunov$^{\rm 1}$\footnotemark[2]\footnotetext[2]
{E-mail: ivan@phys.tsu.ru}}
and S.L. Lyakhovich$^{\rm 1,2}$\footnotemark[3]\footnotetext[3]
{E-mail: sll@phys.tsu.ru}\\[0.2cm]
\footnotesize{ ${}^1${\it  Department of Physics, Tomsk State University}\\
Lenin Ave. 36, Tomsk, 634050 Russia} \\
\footnotesize{${}^2${\it International Centre for Theoretical Physics}\\
P.O. Box 586, 3410 Trieste, Italy}
\end{center}
\vspace{0.6cm}

\begin{abstract}
The first quantized theory of N=2, D=3 massive superparticles with arbitrary
fixed central charge and (half)integer or fractional superspin is
constructed. The quantum states are realized on the fields carrying a
finite dimensional, or a unitary infinite dimensional representation of the
supergroups $\rm OSp(2|2)$ or $\rm SU(1,1|2)$. The construction originates
from quantization of a classical model of the superparticle we suggest. The
physical phase space of the classical superparticle is embedded
in a symplectic superspace $T^\ast({\rm R}^{1,2})\times{\cal
L}^{1|2}$, where the inner K\"ahler supermanifold $\rm{\cal L}^{1|2}\cong
OSp(2|2)/[U(1)\times U(1)]\cong SU(1,1|2)/[U(2|2)\times U(1)]$ provides the
particle with superspin degrees of freedom. We find the relationship between
Hamiltonian generators of the global Poincar\'e supersymmetry and the
``internal'' $\rm SU(1,1|2)$ one.  Quantization of the superparticle combines
the Berezin quantization on ${\cal L}^{1|2}$ and the conventional Dirac
quantization with respect to space-time degrees of freedom. Surprisingly,
to retain the supersymmetry, quantum corrections are required for the
classical N=2 supercharges as compared to the conventional Berezin method.
These corrections are derived and the Berezin correspondence principle for
${\cal L}^{1|2}$ underlying their origin is verified. The model admits
a smooth contraction to the N=1 supersymmetry in the BPS limit.
\end{abstract}
PACS numbers: 11.30.Pb, 71.10.Pm\\
Keywords: quantization, superparticles, anyons, $\rm OSp(2|2)$.

%\today
\end{titlepage}
\renewcommand{\thefootnote}{\arabic{footnote}}
\setcounter{footnote}{0}
%\newpage\thispagestyle{empty}\tableofcontents
\newpage
\setcounter{page}{1}
%%%%%%%%%%%%%%%%%%%%%%%%%%%%%%%%%%%%%%%%%%%%%%%%%%%%%%%%%%%%%%%%%%%%%%%%%%%
\section[\hspace{4mm}Introduction]
{\normalsize\bf\hspace{-5mm}INTRODUCTION}\label{s0}\hspace*{\parindent}%
%%%%%%%%%%%%%%%%%%%%%%%%%%%%%%%%%%%%%%%%%%%%%%%%%%%%%%%%%%%%%%%%%%%%%%%%%%%
In this paper we construct  N=2, D=1+2 massive spinning superparticle model
and study the symplectic supergeometry behind it.
This supergeometry is compatible to the Berezin quantization method
which is applied to construct the one-particle quantum theory.
The main part of our consideration is based on the observation that
the N=2 superextension of D=3 spinning particle results in the classical
model which possesses simultaneously Poincar\'e supersymmetry and Lorentz
supersymmetry of the superspin degrees of freedom. This ``double''
supersymmetry can be lifted to the quantum level and we obtain the
realization of N=2, D=3 Poincar\'e supermultiplet on the fields carrying
an irreducible representation of the supergroup $\rm SU(1,1|2)$ (``Lorentz
supergroup'' whose even part is $\rm SO^\uparrow(1,2)\times U(2)\times
central\ charge$). A nonlinear mutual involvement of the Hamiltonian
generators of two supersymmetries requires the careful geometric
quantization of the superparticle. At first, we try to explain the most
important motivations of the problem.

In the hierarchy of all known entities, the particles living in
three-dimensional space-time stand out mostly due to a possibility of
fractional spin and statistics (anyons). Anyon excitations are actually
presented in some planar physics phenomena \cite{Laugh,Wilczekbook} and the
relevant theoretical concept has both topological \cite{LeiMyr,Wilc,Goldin}
and group-theoretical \cite{JackNair,Forte,SorVol} grounds. It is well known
that in the field theory fractional statistics originates usually from a
coupling of the matter fields to the gauge field with the Chern-Simons mass
term~\cite{Chern-Sim}. The supersymmetric extension of this approach
\cite{HlousSpect} implies a direct interaction between anyon excitations.

The group-theoretical methods may give an alternative way to understand
the anyon concept. One can start from the mechanical model of D=3 spinning
particle, whose quantization leads to the one-particle quantum mechanics for
the fractional spin state
\cite{Balach,Plyush,SorVol,CorPly,Ghosh95,GKL1,Ners}. It is established
that D=3 spinning particles possess the following remarkable features:
(i)~the spinning particle carries as many physical degrees of freedom as a
spinless one; (ii)~there is the so-called {\it canonical\/} model
\cite{Balach} of spinning particle, which implies a deformation of the
canonical symplectic structure of spinless particle by the use of {the Dirac
monopole two-form\/}, without extension of the phase space introducing any
``spinning'' variables; (iii)
it is promising feature of the canonical model to be
adapted for the construction of consistent couplings of the particle
to external fields \cite{ChouNairPol,Chou,Ghosh95,GhoshMukh,GKL1} and
self-interaction of anyons \cite{BenGhosh,Ghosh97.1,Ghosh97.2}. In higher
dimensions, the interaction problem for spinning particles becomes more
involved, although some progress has recently been  achieved there as well
\cite{LyakSegShar,LShSh98}; (iv) the anyon wave equations may be
formulated in analogy with the ones for bosons and fermions.  An essential
difference is that the fractional spin, in contrast to the (half)integer,
is naturally described in terms of infinite component fields carrying
infinite dimensional representations of the universal covering group
$\overline{{\rm SO}^\uparrow(1,2)}\cong\overline{{\rm SU}(1,1)}$;
(v) representations of fractional spin are multivalued.

There is no consistent quantum field theory of anyons up to now, nevertheless
the Chern-Simons and group-theoretical constructions are deemed to lead to a
unified consistent theory. In this regard, it would be interesting to
understand, how the supersymmetry may be included into a group-theoretical
description of anyons in terms of the infinite component fields.

Another reason to investigate the D=3 superparticle is the exceptional fact
that not only the Poincar\'e supersymmetry is possible in 1+2 dimensions, but
the Lorentz one is too. The Lorentz group $\rm SO^\uparrow(1,2)$
coincides with the D=2 anti-de Sitter group, the latter admits the
superextension regardless of specific space-time dimension. Although the
Lorentz and the Poincar\'e supersymmetries are not compatible with each other,
surprisingly, we will show that the Lorentz supersymmetry of D=3 spinning
superparticle (which is invariant by construction with respect to the global
Poincar\'e SUSY transformations) manifests itself as a hidden
supersymmetry of internal degrees of freedom associated to the particle
superspin and to the underlying superextended monopole-like symplectic
structure.

The hidden $\rm OSp(2|2)$ supersymmetry of N=1 superanyons has been
found in Ref.~\cite{GKL2} where the respective model is
constructed.
The presence of the $\rm OSp(2|2)$ supersymmetry already in the
classical mechanics appears to be crucial for a consistent first quantization
of N=1, D=3 superanyon. As a result, one obtains in quantum theory the
realization of the N=1 Poincar\'e supermultiplet on the fields carrying an
atypical unitary infinite dimensional representation of the $\rm
OSp(2|2)$~\cite{GKL2}. It is a direct N=1 superextension of description in
terms of infinite dimensional unitary representation of the D=3 Lorentz
group~\cite{JackNair,Forte,SorVol} or the ones of the deformed Heisenberg
algebra~\cite{Ply1}. We argued in this manner the relevance of the
group-theoretical approach for N=1 supersymmetric anyons. In this paper we
suggest a nontrivial generalization of this construction to the case of D=3,
N=2 massive spinning superparticle with arbitrary fixed central charge.

We construct a superparticle model, which gives N=2
superextension of the canonical description of the D=3 spinning particle
mentioned above. It is essential for our consideration that the Hamiltonian
formalism of the canonical model may be built either in terms of the minimal
phase space, or in an extended phase space restricted by
constraints~\cite{CorPly,GKL1,Ners,GKL2}. In both cases the reduced phase
space could be thought of as a space of motion of a Souriau's ``elementary
system''~\cite{Sour}.

A general concept of elementary physical systems, including spinning
particles and superparticles, is based on the so-called
Kostant-Souriau-Kirillov
construction \cite{Kost,Sour,Kiril}. The idea of the KSK construction is to
identify the {\it physical phase space (= space of motion)\/} of any
elementary system with a {\it coadjoint orbit} $\cal O$ of the symmetry
group~$G$. The symplectic action $G$ on $\cal O$ (classical mechanics) lifts
to a representation of the group in a space of  functions $\cal H$ on the
classical manifold (prequantization). Then the quantization problem
reduces to an appropriate choice of polarization,
that is a global Lagrangian section in
$T({\cal O})$ being invariant under the action of the symmetry group.

In the special case  of K\"ahler homogeneous spaces perfect
results can be achieved
in the framework of the Berezin quantization method~\cite{Berezin1,Berezin2},
which implies one-to-one correspondence between the phase-space functions
(covariant Berezin symbols) and linear operators in a Hilbert space. The
latter is realized by holomorphic sections, because the K\"ahler
homogeneous manifold admits a natural complex polarization~\cite{Wood}.
Moreover, the multiplication of the operators in the Hilbert space induces a
noncommutative binary \mbox{$\ast$-operation} for the covariant Berezin
symbols and a correspondence principle can be proved~\cite{Berezin1,Berezin2}.

Physically speaking, it would not  always be satisfactory to describe
elementary systems in terms of the coadjoint orbits. In particular, the
dynamics of relativistic particles and superparticles is usually supposed
to evolve in a fibre bundle $\cal M$ over
a {\it space-time manifold\/} that is crucial for the interaction problem.
Thus, the coadjoint orbit of the spinning (super)particle arises from
embedding into evolution (super)space. The projection $\pi:{\cal
M}\to{\cal O}_G\,$, where $G$ is a Poincar\'e (super)group, generates
$G$-invariant constraints and gauge symmetries in $\cal M$. The
construction of interactions, being consistent with the gauge symmetries,
and the quantization problem for $\pi$ provide a subject of current
interest in the problems of spinning particle and superparticle
models~\cite{Balach,LyakSegShar,LShSh98,MarnMart,Frydr,KuzLyakSeg}.

Concerning the D=3 spinning particle, we know~\cite{GKL1,GKL2} that the
quantization problem for the canonical model is naturally solved by means of an
embedding of the maximal (four-dimensional) coadjoint orbit of the group ${\rm
ISO}^\uparrow (1,2)$ into eight dimensional
phase space (that is {\it extended phase space}) ${\cal M}^8\cong
T^\ast({\rm R}^{1,2})\times {\cal L}$. Here ${\cal L}\cong\rm SU(1,1)/U(1)$
is a Lobachevsky plane and the character $\cong$ denotes a symplectomorphism.
The projection $\pi:{\cal M}^8\to{\cal O}_{m,s}$ onto co-orbit ${\cal
O}_{m,s}$ of the particle of mass $m$ and spin $s$ is provided by the
constraints.  The auxiliary variables parametrizing $\cal L$ are used to
describe the particle spin. One can interpret the (holomorphic)
automorphisms of the Lobachevskian K\"ahler metric as a hidden symmetry of
the internal particle's structure, which is related to the spin. We observed
in Refs.~\cite{GKL1,GKL2} that the quantization of anyon could be achieved as
a compromise of the conventional Dirac quantization on $T^\ast({\rm
R}^{1,2})$ and of the geometric quantization in the Lobachevsky plane.
Constraints of the classical mechanics are converted into wave equations of
anyon according to the Dirac prescriptions.

The starting point of this paper is a mechanical model of N=2
superparticle with arbitrary fixed mass $m>0$, superspin $s\neq0$ and central
charge ${\cal Z}\equiv mb$, ${|b|\leq1}$ briefly announced before
\cite{GL98}.  For this elementary system the maximal coadjoint orbit
${\cal O}_{m,s,b}$ of real dimension $4/4$ is related to the case $|b|<1$.
In our model, this orbit appears  embedded into $8/4$-dimensional
extended phase superspace ${\cal M}^{8|4}$ of a special geometry: ${\cal
M}^{8|4}\cong T^\ast({\rm R}^{1,2})\times {\cal L}^{1|2}$, where $\rm
{\cal L}^{1|2} =SU(1,1|2)/[U(2|2)\times U(1)]\cong OSp(2|2)/[U(1)\times
U(1)]$ is an atypical K\"ahler coadjoint orbit of the supergroup $\rm
SU(1,1|2)$ and the typical one of  $\rm OSp(2|2)$. The inner supermanifold
${\cal L}^{1|2}$, providing the particle model with a nonzero superspin,
was studied originally in Refs.~\cite{Balant,Grad} in relation to $\rm
OSp(2|2)$ supercoherent states and called {\it N=2 superunit disc\/}.  The
projection of ${\cal M}^{8|4}$ onto physical subspace follows similarly to
the nonsupersymmetric model. In fact, introducing the supersymmetry for
D=3 particle, we need to superextend only the inner submanifold $\cal L$ of
the extended phase space. The extended phase superspace ${\cal
M}^{8|4}\cong T^\ast({\rm R}^{1,2})\times {\cal L}^{1|2}$ carries ``double
supersymmetry'':  one is related to the Poincar\'e supergroup and acts on the
associated co-orbit~${\cal O}_{m,s,b}\subset{\cal M}^{8|4}$, another one
lives in the inner subsupermanifold~${\cal L}^{1|2}$. Moreover, the model
allows an extended hidden N=4 supersymmetry with special values of the
central charges saturating the BPS bound.

We will quantize the theory similarly to quantization of the canonical
model of the particle on ${\cal M}^8$ \cite{GKL1,GKL2}. Specifically, we
combine the geometric quantization in the inner subsupermanifold ${\cal
L}^{1|2}$ for the internal $\rm SU(1,1|2)$ supersymmetry and the canonical
Dirac quantization in $T^\ast({\rm R}^{1,2})$.

This quantization scheme implies from the outset that the mentioned ``double
supersymmetry'' must survive in the quantum theory. The crucial
point is to express the Hamiltonian generators of the Poincar\'e
supersymmetry in ${\cal M}^{8|4}$ in terms of the ones of internal $\rm
SU(1,1|2)$ supersymmetry (as well as of space-time coordinates and momenta).
These expressions appear to be {\it nonlinear}. As a consequence, some
renormalization of the Poincar\'e supergenerators should be required for the
closure of the Poincar\'e supersymmetry algebra. Roughly speaking, the
corrections to generators could be treated as a manifestation of the ordering
ambiguity for operators in quantum theory. We will see, that the origin of
the corrections may  also be clarified from the viewpoint of the Berezin
quantization in ${\cal L}^{1|2}$ and the underlying correspondence principle.
However, the Berezin method itself does not provide a regular technique of
deriving the closing corrections which have to recover the
representation of the Poincar\'e superalgebra in quantum theory. Moreover,
it is unclear a priori whether the consistent corrections exist at all.
Surprisingly, the problem is solved if a simple ansatz is taken for the
renormalized Poincar\'e generators. Then the closing corrections, which
appear in the order of ${\cal O}(s^{-2})$, can be {\it exactly}
calculated.

We arrive eventually to the realization of the unitary representation of
N=2, D=3 supermultiplet on the fields carrying atypical irreps of the
supergroup $\rm SU(1,1|2)$ and the typical ones of the subsupergroup $\rm
OSp(2|2)$. These irreps are certainly infinite dimensional for the case of
fractional superspin, but for the habitual case of (half)integer superspin
they may be chosen to be finite dimensional.

The model of N=2 superparticle reduces to the one of N=1 superparticle in the
Bogomol'ny-Prassad-Sommerfield (BPS) limit for central charge, when $|b|=1$.
One can trace the BPS limit both at the classical and quantum levels.
Classically, it corresponds to the degenerate coadjoint orbit of D=3,
N=2 superparticle of dimension~$4/2$.  When $|b|=1$, the extended phase
superspace becomes degenerate and reduces to ${\cal M}^{8|2}\cong
T^\ast({\rm R}^{1,2})\times {\cal L}^{1|1}$ with inner supermanifold
${\cal L}^{1|1}\cong\rm OSp(2|2)/U(1|1)\cong OSp(1|2)/U(1)$.  ${\cal
M}^{8|2}$ is exactly the extended phase superspace of N=1
superanyon~\cite{GKL2}. In this exceptional case, the generators of N=1
Poincar\'e supersymmetry and the internal $\rm OSp(2|2)$ one are {\it
linearly} expressible to one another.  Thus, the geometric quantization
immediately gives the quantum theory of N=1 superparticle, without extra
constructions and corrections. In particular, we don't need the detailed
Berezin correspondence principle for ${\cal L}^{1|1}$.

The geometric quantization in the $\rm OSp(2|2)$ coadjoint orbits was
constructed in Refs.~\cite{Balant,Grad} and we follow these results. At the
same time we have to clarify two important points, which have seemingly been
unknown. First, we found out that the K\"ahler geometry of the regular
co-orbit ${\cal L}^{1|2}$ admits the symplectic holomorphic action of the
supergroup $\rm SU(1,1|2)$, which is larger than the supergroup $\rm
OSp(2|2)$ in itself. We construct the geometric quantization on ${\cal
L}^{1|2}$ provided for this extended supersymmetry supergroup. Secondly,
we perform Berezin quantization for ${\cal L}^{1|2}$ to establish a
correspondence principle and to explain the origin of quantum corrections to
the N=2 Poincar\'e supercharges in ${\cal M}^{8|4}$.

The paper is organized as follows. In Sec.~II we recall briefly the canonical
model of D=3 spinning particle in terms of the minimal and extended phase
spaces. Specifically, we focus at symplectic structure and symmetries
of the minimal and extended spaces.

Then we are going to construct the superextension of the canonical model.
The classical
mechanics of N=2, D=3 massive spinning superparticle with arbitrary central
charge is considered in Sec.~III{}. Starting from a first order Lagrangian we
study the supergeometry of the phase superspace and  identify it with
${\cal M}^{8|4}=T^\ast({\rm R}^{1,2})\times{\cal L}^{1|2}$.
We construct explicitly the embeddings of the N=2 Poincar\'e and Lorentz
supergroup's coadjoint orbits into ${\cal M}^{8|4}$ and find out the
Hamiltonian generators of corresponding supersymmetries. The relation,
being crucial for quantization, is established between the N=2 Poincar\'e
and $\rm SU(1,1|2)$ Hamiltonian generators. We  also reveal a degenerate
N=4 supersymmetry in the model and a special case of degenerate co-orbits,
which appear in the BPS limit.  Furthermore, a reduction of the model with
respect to a part of constraints is shown to lead to a minimal
$6/4$-dimensional phase superspace with superextended symplectic
monopole-like structure and with the mass-shell condition to be the only
constraint. We obtain, in particular, N=2 superextension of the Dirac
monopole two-form, which supplies the particle with superspin.

In Sec.~IV we suggest a quantization procedure for the classical mechanics
constructed in Sec.~III{}. At first the Berezin quantization is considered
on the regular $\rm OSp(2|2)$ coadjoint orbit. In particular, we construct
the correspondence between symbols and operators on ${\cal L}^{1|2}$ and
prove the underlying correspondence principle. Then these results are applied
to the consistent quantization of D=3, N=2 superparticle, which is the final
object of construction.

The summary and a general outlook are given in Sec.~V{}. Finally, the Appendix
contains the calculation of N=2 Poincar\'e supercharge's quantum
anticommutator. The calculation provides manifest verification of
consistency of the renormalization procedure for the Poincar\'e supercharges.

%%%%%%%%%%%%%%%%%%%%%%%%%%%%%%%%%%%%%%%%%%%%%%%%%%%%%%%%%%%%%%%%%%%%%%%%%%%%%
\section[\hspace*{2mm} Minimal and extended phase spaces of spinning
particle]{\normalsize\bf\hspace{-6.5mm} MINIMAL AND
EXTENDED PHASE SPACES \protect\\ \hspace*{-5mm}OF A CANONICAL MODEL OF
SPINNING PARTICLE}\label{s1}%%%%%%%%%%%%%%%%%%%%%%%%%%%%%%%%%%%%%%%%%%%%%%%%%
\hspace*{\parindent}\setcounter{equation}{0}%%%%%%%%%%%%%%%%%%%%%%%%%%%%%%%%%
%%%%%%%%%%%%%%%%%%%%%%%%%%%%%%%%%%%%%%%%%%%%%%%%%%%%%%%%%%%%%%%%%%%%%%%%%%%%%
First consider the nonsupersymmetric canonical model of the particle (various
formulations see~\cite{Balach,JackNair,Forte,CorPly,Ners}), which serves
as an initial subject for further generalizations. The particle lives
originally on six-dimensional phase space ${\cal M}^6$ with a symplectic
two-form\footnote{We use Latin letters to denote D=3 Lorentz vectors and
Greek letters for the $\rm SU(1,1)$ spinors; Minkowski metric is chosen to be
$\eta_{ab}={\rm diag}(-1,1,1)$, totally antisymmetric tensor is normalized by
condition $\epsilon_{012}=-\epsilon^{012}=1$; the spinor indices are raised
and lowered with the use of the spinor metric
$\epsilon^{\alpha\beta}=-\epsilon^{\beta\alpha}= -\epsilon_{\alpha\beta}$
$(\alpha, \beta=0, 1)$, $\epsilon^{01}=-1$ by the rule
$\psi_{\alpha}=\epsilon_{\alpha\beta}\psi^{\beta}, \psi^{\alpha}=
\epsilon^{\alpha\beta}\psi_{\beta}$.}
\begin{equation}
\Omega_s=-{\rm d}x^a\wedge{\rm d}p_a+\Omega_{\rm m}\qquad
\Omega_{\rm m}=\frac{s}2
\frac{\epsilon^{abc}p_a{\rm d}p_b\wedge{\rm d}p_c}{(-p^2)^{3/2}}\qquad
(p^2<0)\,,
\label{2.1}
\end{equation}
where $\Omega_{\rm m}$ is known as the Dirac monopole form. The
Poincar\'e transformations are generated by the following functions
\begin{equation}
{\cal P}_a=p_a\qquad{\cal J}_a
=\epsilon_{abc}x^bp^c-s\frac{p_a}{(-p^2)\lefteqn{{}^{1/2}}}\qquad ,
\label{2.2}
\end{equation}
which constitute D=3 Poincar\'e algebra with respect to Poisson
brackets (PB's)
\begin{equation}
\{{\cal P}_a\,,\,{\cal P}_b\}=0\qquad
\{{\cal J}_a\,,\,{\cal P}_b\}=\epsilon_{abc}{\cal P}^c\qquad
\{{\cal J}_a\,,\,{\cal J}_b\}=\epsilon_{abc}{\cal J}^c\,.\label{2.3}
\end{equation}
The fundamental PB's read
\begin{equation}
\{x^a\,,\,x^b\}=s\frac{\epsilon^{abc}p_c}{(-p^2)^{3/2}}\qquad
\{x^a\,,\,p_b\}=\delta^a{}_b\qquad
\{p_a\,,\,p_b\}=0\,.\label{2.4}
\end{equation}
The last two PB's mean that $x^a$ and $p_a$ transform as
coordinates and momenta by Poincar\'e translations. Moreover, they are
Lorentz vectors because of $\{{\cal J}_a\,,\,x_b\}=\epsilon_{abc}x^c$ and
$\{{\cal J}_a\,,\,p_b\}=\epsilon_{abc}p^c$.

Let us assume that the particle dynamics on ${\cal M}^6$ is governed by the
mass shell {\it constraint}
\begin{equation}
p^2+m^2=0\,,\label{2.5}
\end{equation}
whereas the canonical Hamilton function is identically zero. On the mass
shell, the Casimir functions of the enveloping Poincar\'e algebra are
identically conserved: ${\cal P}^2=-m^2$, $({\cal P},{\cal J})=ms$.  We
conclude that D=3 particle of mass $m$, spin $s$ and energy sign $p^0/|p^0|$
lives on mass shell. From now on, we take a further restriction $p^0>0$
bearing in mind the supersymmetric theory, when the energy is positive
essentially. The mass shell constraint generates the
reparametrization (gauge) invariance for every world line of the particle.
The set of world lines, being considered modulo to the gauge equivalence,
is named the particle history space, the latter is isomorphic to the physical
state space~${\cal O}_{m,s}$ of the spinning particle. The reduced symplectic
manifold~${\cal O}_{m,s}$ is symplectomorphic to the maximal {\it coadjoint
orbit}~\cite{Sour,Kiril,Wood} of the D=3 Poincar\'e group.

There is a standard way to extend the canonical model to the Poincar\'e
supersymmetry. One may substitute ${\rm d}x^a\to {\rm d}
x^a-i(\gamma^a)_{\alpha \beta}\theta^{\alpha\,I}{\rm d}\theta^{\beta\,I}$ in
Eq.\ (\ref{2.1}) introducing real Grassmann variables
$\theta^{\alpha\,I},\,I=1,\dots,N$.  The resulting symplectic superform
appears to be invariant under the $N$-extended Poincar\'e supergroup without
central charges. One may further generate central charges introducing some
Wess-Zumino type terms~\cite{AzcLuk,Frydr} in Eq.\ (\ref{2.1}). Then, imposing
the mass shell constraint (\ref{2.5}), one may build the classical model of
D=3 superparticle of mass $m$, superspin $s$ and arbitrary fixed central
charges in the $6/2N$-dimensional phase superspace. However, it is hardly
possible to conceive satisfactory quantization of this model.

Even for the canonical model without supersymmetry the realization of
the coordinate operators ${\widehat x}{}^a$ is a nontrivial problem
accounting for the complicated form of the first Poisson bracket in Eqs.\
(\ref{2.4}). A detailed analysis of Ref.~\cite{CorPly} shows that the
manifest covariance of the canonical model, being formulated in terms of the
``minimal'' phase space ${\cal M}^6$, is inevitably lost in quantum
theory.  The superextension of the canonical model makes the Poisson
brackets, being quantized, much more complicated. In fact, the
quantization problem in the reduced nonlinear phase superspace is not
solved even for spinless D=3 superparticle. Thus we will reformulate from
the outset the canonical model in an ``extended'' phase space, where a
hidden symmetry of spinning particle becomes transparent and gives an
efficient method for quantization making use of this symmetry. Moreover,
the construction will be appropriate for intriguing superextension.

An adapted reformulation of the canonical model is suggested in
Refs.~ \cite{GKL1,GKL2}. We observe, that the monopole two-form
$\Omega_{\rm m}$ in Eq.\ (\ref{2.1}) is nothing else but the K\"ahler two
form on the mass hyperboloid (\ref{2.5}), which gives the realization of
the Lobachevsky plane ${\cal L}$. It will be convenient to make use of
another realization of ${\cal L}\cong\{z\in {\rm C}^1, |z|<1\}$ by an open
unit disc of complex plane ${\rm C}^1$. We rewrite the symplectic
two-form~(\ref{2.1}) as follows
\begin{equation}
\Omega_s=-{\rm d}x^a\wedge{\rm d} p_a+\Omega_{\cal L}\qquad\Omega_{\cal L}=
-2is\frac{{\rm d}z\wedge{\rm d}\bar z}{(1-z\bar z)^2}\,,
\label{2.6}
\end{equation}
where (recall that $p^2<0$ and we have taken $p^0>0$)
\begin{equation}
p^a=\sqrt{-p^2}n^a\qquad n^a=\left(\frac{1+z\bar z}{1-z\bar z},
-\frac{z+\bar z}{1-z\bar z},i\frac{\bar z-z}{1-z\bar z}\right)\quad
n^2\equiv-1\,.
\label{2.7}
\end{equation}
The unit timelike Lorentz vector $n^a$ parametrizes the points of the
Lobachevsky plane.

Let us look at Eq.\ (\ref{2.6}) from a different viewpoint. Consider a
new phase space ${\cal M}^8\cong T^\ast({\rm R}^{1,2})\times{\cal L}$
with a symplectic two-form (\ref{2.6}) and an elementary system on
${\cal M}^8$, whose dynamics is subjected by three constraints
\begin{equation}
p^a=mn^a
\label{2.8}
\end{equation}
Apparently these constraints project the extended phase space ${\cal M}^8$
into the same coadjoint orbit as the mass shell constraint (\ref{2.5}) does for
${\cal M}^6$. Alternatively, one can solve explicitly only two
constraints $p^a=\sqrt{-p^2}n^a$ providing the reduction
${\pi_1:{\cal M}^8}\to{\cal M}^6$ of extended phase space to the minimal
one. In other words, we have constructed the sequence of embeddings ${\cal
O}_{m,s}\subset {\cal M}^6\subset{\cal M}^8$. Hence we get an equivalent
description of D=3 spinning particle in terms of the extended phase space
$T^\ast({\rm R}^{1,2})\times{\cal L}$.  The Hamiltonian generators of the
canonical Poincar\'e transformations in ${\cal M}^8$ read
\begin{equation}
{\cal P}_a=p_a\qquad {\cal J}_a=\epsilon_{abc}x^bp^c+J_a\,,
\label{2.9}
\end{equation}
where the spin vector $J_a$ is expressed in terms of the `inner' space
$\cal L$:
\begin{equation}
J_a=-sn_a \, .
\label{2.10}
\end{equation}
The Hamiltonians (\ref{2.9}) generate the Poincar\'e algebra with respect
to PB in ${\cal M}^8$, whereas the spin generators (\ref{2.10}) span internal
Lorentz algebra related to the (holomorphic) automorphism group of the
Lobachevsky plane. The latter group can be recognized as a hidden symmetry of
the internal structure of spinning particle. Although this concept looks
may seem artificial at the moment, below, we will observe
essentially nontrivial superextension of the hidden symmetry.

The Poincar\'e Casimir functions are identically conserved owing to
constraints~(\ref{2.8})
\begin{equation}
p^2+m^2=0\qquad (p,J)-ms=0\,.\label{2.11}
\end{equation}
A crucial detail is that the equations (\ref{2.11}) define
the same surface in the extended phase space as the constraints
(\ref{2.8}) do.

The quantization of the model in ${\cal M}^8$ is almost transparent. We
can combine the canonical Dirac quantization in $T^\ast({\rm R}^{1,2})$
and the Berezin quantization in the Lobachevsky plane~\cite{GKL2}.
Constraints (\ref{2.11}) will be imposed in Hilbert space to separate the
one-particle states.

Finally, write down the Lagrangian of the theory.
One may choose the action functional as an integrand of the one-form $\Theta$,
where ${\rm d}\Theta=\Omega_s+V$ and $V$ vanishes on shell. Let us take
\begin{equation}
S=\int\Theta \, , \qquad
\Theta=p_a{\rm d}x^a
+is\frac{\bar z{\rm d}z -
z{\rm d}\bar z}{1-z\bar z}\equiv p_a{\rm d}x^a+\Sigma_{\cal L}\, ,
\qquad
{\rm d}\Sigma_{\cal L}=\Omega_{\cal L}\,.
\label{2.12}
\end{equation}
It is implied here that the virtual paths lay in the constraint
surface (\ref{2.8}).  Excluding the momenta accounting for
constraints (\ref{2.8}) and making pull back of $\Theta$, one obtains the
action functional
\begin{equation}
S=\int\limits_{\tau_1}^{\tau_2}L\,{\rm
d}\tau\qquad L=m(\dot x,n) +is\frac{\bar z{\dot z} -z\dot{\bar z}}{1-z\bar z}
\label{2.13}
\end{equation}
with the first order Lagrangian being invariant under reparametrizations.
Notice that the Lagrangian is also strongly invariant under translations
and spatial rotations, whereas the Lorentz boosts change it by a total
derivative.

%%%%%%%%%%%%%%%%%%%%%%%%%%%%%%%%%%%%%%%%%%%%%%%%%%%%%%%%%%%%%%%%%%%%%%%%
\section[\hspace*{2mm} Classical model of D=3 spinning superparticle]
{\normalsize\bf
\hspace{-6.5mm}CLASSICAL MODEL OF $\normalsize\bf D=3$ SPINNING SUPERPARTICLE}
\label{s3}\nopagebreak\setcounter{equation}{0}%%%%%%%%%%%%%%%%%%%%%%%%%
%%%%%%%%%%%%%%%%%%%%%%%%%%%%%%%%%%%%%%%%%%%%%%%%%%%%%%%%%%%%%%%%%%%%%%%%
\subsection[A first order Lagrangian]
{\normalsize\bf\hspace{-4mm} A first order Lagrangian}
\label{sbs3.1}\hspace*{\parindent}\nopagebreak%%%%%%%%%%%%%%%%%%%%%%%
%%%%%%%%%%%%%%%%%%%%%%%%%%%%%%%%%%%%%%%%%%%%%%%%%%%%%%%%%%%%%%%%%%%%%%%
Introduce a N=2 superextension of the Lagrangian (\ref{2.13})
providing both the superpoincar\'e invariance of the theory and
other hidden supersymmetry as well. Introducing a pair of Majorana
anticommuting spinors $\theta^{\alpha\,I}=(\theta^\alpha,\chi^\alpha), I=1,2$
we suggest
\begin{equation}
L=m(\Pi,n)+mb(\theta_\alpha\dot\chi^\alpha-\chi_\alpha\dot\theta^\alpha)-
mb\theta^\alpha n_{\alpha\gamma}\dot n^{\gamma}{}_\beta\chi^\beta
+is\frac{\bar z{\dot z} -z\dot{\bar z}}{1-z\bar z}\,,
\label{3.1}
\end{equation}
where $m,b,s$ are real parameters, $n^a$ is a unit Lorentz vector
in the  Lobachevsky plane, being defined by Eq.~(\ref{2.7}) and
\begin{displaymath}
n_{\alpha\beta}\equiv n^a\gamma_{a\,\alpha\beta}\qquad\qquad
\Pi^a=\dot x^a-i\gamma^a_{\alpha\beta}(\theta^\alpha\dot\theta^\beta
+\chi^\alpha\dot\chi^\beta)\,.
\end{displaymath}
The three dimensional Dirac matrices $\gamma_a$ are chosen in the
form\footnote{One may wonder, why the $\gamma$-matrices are not Hermitian.
It is instructive to note that the reality condition for $\rm SU(1,1)$
spinor formalism is not trivial, as for isomorphic $\rm SL(2,R)$ ones.
For any $g\in\rm SU(1,1)$ the complex conjugation reads $\bar g=cgc$, where
$c=c^{-1}=\rm antidiag(-1,-1)$. The matrices $c\gamma^a$ are truly Hermitian.
The covariant Majorana (reality) condition looks like
\begin{equation}
c\bar\psi=\psi\label{Major}
\end{equation}
for two-component $\rm SU(1,1)$-spinor $\psi$.\label{ftn2}}
\begin{displaymath}
\begin{array}{c}
(\gamma_0)_{\alpha\beta}=
\left( \begin{array}{cc}
            0 & 1 \\
            1 & 0
       \end{array}\right)\quad
(\gamma_1)_{\alpha\beta}=
\left( \begin{array}{cc}
            1 & 0 \\
            0 & 1
       \end{array}\right)\quad
(\gamma_2)_{\alpha\beta}=
\left( \begin{array}{cc}
           -i  & 0 \\
            0 & i
       \end{array}\right)\\
(\gamma_a)_{\alpha\gamma}(\gamma_b)^\gamma{}_\beta=
i\epsilon_{abc}(\gamma^c)_{\alpha\beta}-\eta_{ab}\epsilon_{\alpha\beta}\,.
\end{array}
\end{displaymath}
The first term in the Lagrangian
(\ref{3.1}) is a conventional superextension of
the respective expression in Eq.~(\ref{2.13}) and the second addend
represents the Wess-Zumino type term generating the central charge for the
supersymmetry \cite{AzcLuk}.  At last, the third term accounts for the
specific of D=3 spinning superparticle model.  Owing to this addend the
supertranslations, underlying Poincar\'e supersymmetry of the Lagrangian,
read rather unusual:
\begin{equation}
\begin{array}{llll}
\displaystyle
\delta_\epsilon x^a=i\gamma^a_{\alpha\beta}\epsilon^\alpha\theta^\beta+
ib\epsilon^{abc} n_b \gamma_{c\,\alpha\beta}\epsilon^\alpha\chi^\beta+
bn^a\epsilon^\alpha\chi_\alpha &
\displaystyle
\delta_\epsilon \theta^\alpha=\epsilon^\alpha &
\displaystyle
\delta_\epsilon \chi^\alpha=0 &
\displaystyle
\delta_\epsilon z=0 \\
\displaystyle
\delta_\eta x^a=i\gamma^a_{\alpha\beta}\eta^\alpha\chi^\beta-
ib\epsilon^{abc} n_b \gamma_{c\,\alpha\beta}\eta^\alpha\theta^\beta-
bn^a\eta^\alpha\theta_\alpha &
\displaystyle
\delta_\eta \theta^\alpha=0 &
\displaystyle
\delta_\eta \chi^\alpha=\eta^\alpha &
\displaystyle
\delta_\eta z=0\,.
\end{array}
\label{3.2}
\end{equation}
Here $\epsilon^\alpha,\eta^\alpha$ are odd real parameters. For
completeness, expose also the even infinitesimal Poincar\'e
transformations and
$\rm U(1)$ transformations as well
\begin{equation}
\begin{array}{llll}
\displaystyle
\delta_\omega x^a=\epsilon^{abc}\omega_bx_c&
\displaystyle
\delta_\omega \theta^\alpha=-\frac{i}2 \omega^a\gamma_a{}^\alpha{}_\beta
\theta^\beta&
\displaystyle
\delta_\omega \chi^\alpha=-\frac{i}2
\omega^a\gamma_a{}^\alpha{}_\beta\chi^\beta&
\displaystyle
\delta_\omega z=i\omega^a\xi_a \\
\displaystyle
\delta_{\rm f} x^a=f^a&
\displaystyle
\delta_{\rm f} \theta^\alpha=\delta_{\rm f}\chi^\alpha=0 &
\displaystyle
\delta_{\rm f} z=0& \\
\displaystyle
\delta_{\mu} x^a=0&
\displaystyle
\delta_{\mu} \theta^\alpha=-\mu\theta^\alpha &
\displaystyle
\delta_{\mu} \chi^\alpha=\mu\theta^\alpha &
\displaystyle
\delta_{\mu} z=0
\end{array}
\label{3.3}
\end{equation}
with the even real parameters $\omega^a,f^a,\mu$ and the holomorphic object
$\xi_a=-1/2(2z, 1+z^2,i(1-z^2))$. The infinitesimal transformations
(\ref{3.3}), (\ref{3.2}) generate N=2 Poincar\'e superalgebra, which is
discussed in Subsection~3.

%%%%%%%%%%%%%%%%%%%%%%%%%%%%%%%%%%%%%%%%%%%%%%%%%%%%%%%%%%%%%%%%%%%%%%%%
\subsection[Extended phase superspace]{\normalsize\bf
Extended phase superspace}\hspace*{\parindent}%
\rm\label{sbs3.2}\nopagebreak%%%%%%%%%%%%%%%%%%%%%%%%%%%%%%%%%%%%%%%%%%%%
%%%%%%%%%%%%%%%%%%%%%%%%%%%%%%%%%%%%%%%%%%%%%%%%%%%%%%%%%%%%%%%%%%%%%%%%
We show in this subsection that the superparticle being described by the
Lagrangian (\ref{3.1}) lives in a supersymplectic phase space ${\cal
M}^{8|4}$ of very special supergeometry:
${\cal M}^{8|4}\cong T^\ast({\rm R}^{1,2})\times{\cal L}^{1|2}$. Then
we identify ${\cal L}^{1|2}$ with regular (when $|b|<1$) or degenerate
(when $|b|=1$) coadjoint orbit of the $\rm OSp(2|2)$ supergroup.
Having the goal to quantize the theory in ${\cal M}^{8|4}$ we will need for
detailed information about SUSY's and quantization in ${\cal L}^{1|2}$.
The supersymplectic geometry of ${\cal L}^{1|2}$ is considered in
Subsec.~4, while the Berezin quantization will be constructed in
Subsec.~IV.1.

The model (\ref{3.1}) fits naturally into the formulation in symplectic
language. The theory originates from the action functional
\begin{equation}
S=\int\Theta^{\rm SUSY} \qquad \Theta^{\rm SUSY}=p_a{\rm d}x^a+
\Sigma_{{\cal L}^{1|2}}
\label{3.4}
\end{equation}
\begin{eqnarray}&\displaystyle
\Sigma_{{\cal L}^{1|2}}=-imn_{\alpha\beta}\theta^\alpha{\rm d}\theta^\beta
-imn_{\alpha\beta}\chi^\alpha{\rm d}\chi^\beta+mb\theta_\alpha{\rm d}
\chi^\alpha-mb\chi_\alpha{\rm d}\theta^\alpha
&\nonumber\\&\displaystyle\phantom{001}
-2mb\frac{z^\alpha z^\beta\theta_\alpha
\chi_\beta{\rm d}\bar z- \bar z^\alpha\bar z^\beta\theta_\alpha\chi_\beta
{\rm d}z}{(1-z\bar z)^2}+is\frac{\bar z{\rm d}z -z{\rm d}\bar z}{1-z\bar
z}\,,&
\label{3.5}
\end{eqnarray}
where the virtual paths belong  the surface
\begin{equation}
p^a=mn^a\,,
\label{3.6}
\end{equation}
as follows from the definition $p_a=\partial L/\partial\dot x^a$.
Introduce the objects
\begin{equation}
z^\alpha\equiv(1,z)\qquad\qquad\bar z^\alpha\equiv(\bar z,1)\, , \quad
\alpha =0,1 \, ,
\label{3.7}
\end{equation}
that simplifies Eq.~(\ref{3.5}) and many of the forthcoming formulae. In this
section $z^\alpha,\bar z^\alpha$ are used for notation only. Remarkable
origin and transformation properties of the objects $z^\alpha,\bar z^\alpha$
will be considered later in Subsec.~IV.4.

Relation (\ref{3.4}) shows that the particle dynamics is embedded in phase
superspace ${\cal M}^{8|4}\cong T^\ast({\rm R}^{1,2})\times{\cal L}^{1|2}$
with some inner superspace of a real dimension $2/4$ denoted by ${\cal
L}^{1|2}$. The symplectic two-superform in ${\cal M}^{8|4}$ reads
\begin{equation}
\Omega_s^{\rm SUSY}={\rm d}\Theta^{\rm SUSY}=-{\rm d}x^a\wedge{\rm d}p_a+
\Omega_{{\cal L}^{1|2}} \qquad\qquad \Omega_{{\cal L}^{1|2}}
={\rm d}\Sigma_{{\cal L}^{1|2}}
\label{3.8}
\end{equation}
The inner superspace is a N=2 superextension of the Lobachevsky plane.
We show at first that ${\cal L}^{1|2}$ coincides with a coadjoint orbit of
the $\rm OSp(2|2)$ supergroup. Let us introduce new complex Grassmann variables
($m\neq 0, {s\neq 0}$)
\begin{equation}
\begin{array}{l} \displaystyle
\theta=\sqrt\frac{m}{s}(iz^\alpha\chi_\alpha-z^\alpha\theta_\alpha)
\left[1+m\frac{1-b}{4s}(\theta^\alpha\theta_\alpha+\chi^\alpha\chi_\alpha)
\right]\qquad \bar\theta=\overline{(\theta)}\\[1mm] \displaystyle
\chi=\sqrt\frac{m}{s}(iz^\alpha\theta_\alpha-z^\alpha\chi_\alpha)
\left[1+m\frac{1+b}{4s}(\theta^\alpha\theta_\alpha+\chi^\alpha\chi_\alpha)
\right]\qquad \bar\chi=\overline{(\chi)}\,,
\end{array}
\label{3.9}
\end{equation}
which are in one-to-one correspondence with the Majorana spinors
$\theta^\alpha,\chi^\alpha$ used before. It is easy to check that the
symplectic two-superform $\Omega_{{\cal L}^{1|2}}$ of inner superspace reads
in new variables as
\begin{eqnarray}
\displaystyle
\Omega_{{\cal L}^{1|2}}&=& \displaystyle -\frac{is}{2}\left(2
-a_+\frac{1+z\bar z}{1-z\bar z}\theta\bar\theta
-a_-\frac{1+z\bar z}{1-z\bar z}\chi\bar\chi
+a_+a_-\frac{1+2z\bar z}{(1-z\bar z)^2}
\theta\bar\theta\chi\bar\chi\right) \nonumber
\frac{{\rm d}z\wedge{\rm d}\bar z}{(1-z\bar z)^2} \\&&\displaystyle
+\frac{isa_+\bar z\theta}{2}\left(1-\frac{a_-\chi\bar\chi}{1-z\bar z}\right)
\frac{{\rm d}z\wedge{\rm d}\bar\theta}{(1-z\bar z)^2}  \nonumber
+\frac{isa_-\bar z\chi}{2}\left(1-\frac{a_+\theta\bar\theta}{1-z\bar z}\right)
\frac{{\rm d}z\wedge{\rm d}\bar\chi}{(1-z\bar z)^2}\\&&\displaystyle
-\frac{isa_+z\bar\theta}{2}\left(1-\frac{a_-\chi\bar\chi}{1-z\bar z}\right)
\frac{{\rm d}\theta\wedge{\rm d}\bar z}{(1-z\bar z)^2} \nonumber
-\frac{isa_-z\bar\chi}{2}\left(1-\frac{a_+\theta\bar\theta}{1-z\bar z}\right)
\frac{{\rm d}\chi\wedge{\rm d}\bar z}{(1-z\bar z)^2}\\&&\displaystyle
+\frac{isa_+}{2}\left(1-\frac{a_-\chi\bar\chi}{1-z\bar z}\right)
\frac{{\rm d}\theta\wedge{\rm d}\bar\theta}{1-z\bar z}  \label{3.10}
+\frac{isa_-}{2}\left(1-\frac{a_+\theta\bar\theta}{1-z\bar z}\right)
\frac{{\rm d}\chi\wedge{\rm d}\bar\chi}{1-z\bar z}\\&&\displaystyle
+\frac{isa_+a_-\theta\bar\chi}{4}                         \nonumber
\frac{{\rm d}\chi\wedge{\rm d}\bar\theta}{(1-z\bar z)^2}
+\frac{isa_+a_-\chi\bar\theta}{4}
\frac{{\rm d}\theta\wedge{\rm d}\bar\chi}{(1-z\bar z)^2}\\
&&\displaystyle a_+= 1+b\qquad\qquad\qquad a_-= 1-b\,.
\nonumber
\end{eqnarray}
This superform exactly coincides to the one deduced by Gradechi and
Nieto \cite{Grad} in the supercoherent state's approach\footnote{
Explicit form of Eq.\ (\ref{3.10}) depends on the grading conventions
for the exterior superalgebra. We use ${\bf Z}\times{\bf Z_2}$ grading by
analogy with Ref.~\cite{Grad}. The only difference, as compared to
Ref.~\cite{Grad}, is in convention for complex conjugation of the odd
variables.  We take
$\overline{\theta_1\theta_2}=\bar\theta_2\bar\theta_1$, in particular
$\theta\bar\theta$ is a real $c$-number, while in the Ref.~\cite{Grad}
it is an imaginary.}
for the $\rm OSp(2|2)$ coadjoint orbits. $\Omega_{{\cal L}^{1|2}}$
is nondegenerate iff $|b|\neq1$. In the case $|b|<1$, the
supermanifold ${\cal L}^{1|2}$ is the regular $\rm OSp(2|2)$ coadjoint orbit
${\cal L}^{1|2}\cong\rm OSp(2|2)/[U(1)\times U(1)]$ and is called N=2 superunit
disc. The degenerate orbit $\rm OSp(2|2)/U(1|1)$, which is denoted usually by
${\cal L}^{1|1}$ and called N=1 superunit disc, appears when $|b|=1$. The
other possibility $|b|>1$ has no physical significance: neither the Poincar\'e
supersymmetry, nor the internal $\rm OSp(2|2)$ one admit unitary
representations. It is seen from further consideration that the inequality
$|b|>1$ contradicts to the BPS bound.

%%%%%%%%%%%%%%%%%%%%%%%%%%%%%%%%%%%%%%%%%%%%%%%%%%%%%%%%%%%%%%%%%%%%%%%%
\subsection[Observables and the physical subspace]{\normalsize
\bf Observables and the physical subspace}%%%
\hspace*{\parindent}\label{sbs3.3}\nopagebreak%%%%%%%%%%%%%%%%%%%%%%%%%%
%%%%%%%%%%%%%%%%%%%%%%%%%%%%%%%%%%%%%%%%%%%%%%%%%%%%%%%%%%%%%%%%%%%%%%%%
Consider in detail the realization of the Poincar\'e supersymmetry in the
extended phase superspace ${\cal M}^{8|4}\cong T^\ast({\rm
R}^{1,2})\times{\cal L}^{1|2}$. The Poincar\'e supergroup  is realized by
a symplectic action leaving the coadjoint orbit (\ref{3.6}) invariant.
The vector superfields generating the transformations (\ref{3.2}) and
(\ref{3.3}) are related to the corresponding canonical Hamiltonian
generators by
\begin{equation}
X_H\pint \Omega_s^{\rm SUSY}=-(-1)^{\epsilon_H}{\rm d}H\,,\label{3.11}
\end{equation}
where $\epsilon_H$ is the Grassmann parity of the Hamiltonian $H$. Solving
these equations one gets the following Hamiltonian generators (we denote the
generator of isotopic $\rm U(1)$ rotations by $P_3$)
\begin{equation}
\begin{array}{l}\displaystyle
{\cal P}_a=p_a\qquad{\cal J}_a=\epsilon_{abc} x^bp^c-sn_a+\frac12mn_a
(\theta^\alpha\theta_\alpha+\chi^\alpha\chi_\alpha-2ibn_{\alpha\beta}
\theta^\alpha\chi^\beta)\\  \displaystyle
{\cal Q}^1_\alpha=ip_{\alpha\beta}(\theta^\beta-ibn^\beta{}_\gamma\chi^\gamma)
+m(in_{\alpha\beta}\theta^\beta+b\chi_\alpha)\qquad
(p_{\alpha\beta}\equiv p_a\gamma^a_{\alpha\beta})
\\ \displaystyle
{\cal Q}^2_\alpha=ip_{\alpha\beta}(\chi^\beta+ibn^\beta{}_\gamma\theta^\gamma)
+m(in_{\alpha\beta}\chi^\beta-b\theta_\alpha)\\ \displaystyle
P_3=imn_{\alpha\beta}\theta^\alpha\chi^\beta-\frac{mb}2(\theta^\alpha
\theta_\alpha+\chi^\alpha\chi_\alpha)\,.
\end{array}
\label{3.12}
\end{equation}
With respect to Poisson superbrackets on ${\cal M}^{8|4}$ they generate
the following superalgebra
\begin{equation}
\begin{array}{ll}\displaystyle
\{{\cal J}_a\;,\;{\cal J}_b\}=\epsilon_{abc}{\cal J}^c &\displaystyle
\{{\cal J}_a\;,\;{\cal P}_b\}=\epsilon_{abc}{\cal P}^c \qquad
\{{\cal J}_a\;,\;{\cal Q}^I_\alpha\}=-\frac{i}{2}(\gamma_a)_\alpha{}^\beta
{\cal Q}^I_\beta\\ \displaystyle
\{{\cal Q}^I_\alpha\,,\,P_3\}=-\frac12\epsilon^{IJ}{\cal Q}^J_\alpha&
\displaystyle
\{{\cal Q}^I_\alpha\;,\;{\cal Q}^J_\beta\}\approx
-2i\delta^{IJ}p_{\alpha\beta}-2\epsilon^{IJ}
\epsilon_{\alpha\beta}{\cal Z}\qquad  {\cal Z}=mb\,,
\end{array}
\label{3.13}
\end{equation}
the other brackets being equal to zero and $I,J=1,2$,
$\epsilon^{IJ}=-\epsilon^{JI}$, $\epsilon^{01}=1$. We stress that the
latter bracket $\{{\cal Q}^I_\alpha,{\cal Q}^J_\beta\}$
is closed only in a weak sense, that is modulo to
constraints (\ref{3.6}). What we have obtained is N=2, D=3 Poincar\'e
superalgebra with central charge ${\cal Z}=mb$ and isotopic charge $P_3$
acting on the internal indices of supercharges~${\cal Q}^I_\alpha$.

One can easily examine that the mass and the spin Casimir functions of the
superalgebra~(\ref{3.13}) read $C_1\equiv{\cal P}^a{\cal P}_a=p^2$ and
$C_2\equiv{\cal P}^a{\cal J}_a+\frac{1}{8}{\cal Q}^{I\,\alpha}{\cal Q}^I_\alpha
-{\cal Z}P_3=-s(p,n)$. On the constraint surface~(\ref{3.6})
\begin{equation}
p^2+m^2=0\qquad(p,n)+m=0\label{3.14}
\end{equation}
the Casimirs are conserved identically. Eqs.~(\ref{3.14}) and (\ref{3.6})
are completely equivalent to each other, in other words,
they define one and the
same surface in the phase superspace ${\cal M}^{8|4}$. We conclude that
the mechanical model describes N=2, D=3 superparticle of mass $m$, superspin
$s$ and central charge $mb$.

Regular and degenerate cases are essentially distinguished
for the coadjoint orbit, being associated for the superparticle. Since the
massless and spinless particles are not covered in our model, the
Bogomol'nyi-Prassad-Sommerfield bound of central charge (see, for instance,
\cite{sohn}) assumes the only possibility for the degeneracy. The BPS bound
$m\geq|{\cal Z}|$ provides, as is known, consistency of the quantum
theory; the opposite inequality breaks the unitarity.
As we have the goal to construct the quantum theory,
we may restrict the consideration to the case of $|b|\leq 1$.
Furthermore, the limiting point~$|b|=1$ corresponds to the
multiplet-shortening \cite{sohn}. It is the case~$m=|{\cal Z}|$ when the
massive multiplet contains the same number of particles as a massless one.
These massive multiplets are called hypermultiplets. In
the case of~N=2,~D=3 Poincar\'e superalgebra, a massive supermultiplet of
superspin $s$ describes a quartet of particles with spins
$s,s+\frac12,s+\frac12,s+1$ for $m>|{\cal Z}|$ and a doublet $s,s+\frac12$
for $m=|{\cal Z}|$. The shortening of the superparticle multiplet has the
respective origin in the classical mechanics: the number of odd physical
degrees of freedom of the superparticle halfed in the BPS
limit. Let us show that it is the case which is described by our model.

Reducing to the constraints (\ref{3.14}) (or,
equivalently, (\ref{3.6})) we come to the smaller $5/4$-dimensional
phase space ${\cal M}^{5/4}\subset{\cal M}^{8|4}$ with a degenerate
symplectic two-\-super\-form
\begin{equation}
\left.\Omega_s^{\rm SUSY}\right|{}_{p_a=mn_a}
\equiv \Omega_s^{\rm red}=-m{\rm d}x^a\wedge{\rm d}n_a+
\Omega_{{\cal L}^{1|2}} \qquad
{\rm d} n_a\equiv\frac{2\xi_a{\rm d}\bar z+2\bar\xi_a{\rm d}z}{(1-z\bar z)^2}
\,,\label{3.15}
\end{equation}
where $\Omega_{{\cal L}^{1|2}}$ is defined by Eq.\ (\ref{3.10}) and
\begin{equation}
\xi_a= -\frac{1}{2}(\gamma_a )_{\alpha\beta} z^\alpha
z^{\beta}=-\frac{1}{2}(2z,1+z^2,i(z^2-1))\qquad
\bar{\xi}_a=\overline{(\xi_a)}\,.
\label{3.16}
\end{equation}
The kernel of the two-superform (\ref{3.15}) contains obviously the even
one-dimensional null space ${\rm Ker}_0\Omega_s^{red}$, related to the
reparametrization invariance of the world lines. In the coset
superspace~${\cal O}_{m,s,b}={\cal M}^{5|4}/{\rm Ker}_0\Omega_s^{red}$ the
induced symplectic two-superform is nondegenerate when $|b|<1$, the same is
true in~${\cal L}^{1|2}$ for the respective superform~$\Omega_{{\cal
L}^{1|2}}$. Therefore, ${\cal O}_{m,s,b}$, ${\rm dim\,}{\cal
O}_{m,s,b}=4/4\,, |b|<1$ is isomorphic to a regular coadjoint orbit of N=2,
D=3 Poincar\'e supergroup.  We have established both the embedding of the
regular orbit into the original phase superspace and the underlying
projection {$\pi:{\cal M}^{8|4}\to{\cal O}_{m,s,b}$}, provided by
constraints~(\ref{3.14}).

In the BPS limit $|b|=1$ the inner two-superform $\Omega_{{\cal L}^{1|2}}$
generates $0/2$-dimensional null-vector superspace. Thus, the full kernel
${\rm Ker}\,\Omega_s^{red}$ of the symplectic two-superform on ${\cal
M}^{5|4}$ becomes  $1/2$-dimensional if $|b|=1$. The $4/2$-dimensional
coset superspace~${\cal O}_{m,s}={\cal M}^{5|4}/
{\rm Ker}\,\Omega_s^{red}$ corresponds to a degenerate orbit of the N=2
Poincar\'e supergroup. Hence, the number of odd physical
degrees of freedom of~N=2, D=3 superparticle halfed actually in
the BPS limit and we observe an evident classical analogue of the
multiplet-shortening. Some more peculiarities of the BPS limit for
the superparticle model will be discussed in Subsec.~6.

We have described the embedding of coadjoint orbits of the N=2 superparticle
in the phase superspace ${\cal M}^{8|4}\cong T^\ast({\rm
R}^{1,2})\times{\cal L}^{1|2}$. This description should be treated
as a natural superextension of the D=3 spinning particle model with
extended phase space~${\cal M}^8\cong T^\ast({\rm R}^{1,2})\times{\cal
L}$, where the particle spin is realized in terms of the Lobachevsky
plane. One can also construct a different embedding of the superparticle
dynamics generalizing the canonical model with
the minimal six-dimensional phase  space ${\cal M}^6$.  This embedding is
obtained by reduction $\pi_1:
{\cal M}^{8|4}\to{\cal M}^{6|4}$ with respect to two second class
constraints $p_a=\sqrt{-p^2}n_a$ of (\ref{3.6}). If the coordinates in
${\cal M}^{6|4}$ are chosen to be $(x^a,p_a,\theta^{\alpha\,I})$, then the
induced symplectic two-superform reads~as
\begin{eqnarray} \displaystyle
\Omega^{\rm SUSY}_s&=&{\rm d}p_a\wedge{\rm d}x^a+
\left(\frac{s}2+imb\frac{p_{\alpha\beta}\theta^\alpha\chi^\beta}{\sqrt{-p^2}}
\right)\frac{\epsilon^{abc}p_a{\rm d}p_b\wedge{\rm d}p_c}{(-p^2)^{3/2}}
-2imb\epsilon_{\alpha\beta}{\rm d}\theta^\alpha\wedge
{\rm d}\chi^\beta\nonumber
\nonumber \\ &&\displaystyle
-\frac{im}{\sqrt{-p^2}}(\gamma^a)_{\alpha\beta}\left(\Pi_{ab}\theta^\alpha
-b\frac{\epsilon_{abc}p^c}{\sqrt{-p^2}}\chi^\alpha\right){\rm d}p^b\wedge
{\rm d}\theta^\beta \nonumber
-im\frac{p_{\alpha\beta}}{\sqrt{-p^2}}{\rm d}\theta^\alpha\wedge
{\rm d}\theta^\beta
\\ &&\displaystyle
-\frac{im}{\sqrt{-p^2}}(\gamma^a)_{\alpha\beta}\left(\Pi_{ab}\chi^\alpha
+b\frac{\epsilon_{abc}p^c}{\sqrt{-p^2}}\theta^\alpha\right){\rm d}p^b\wedge
{\rm d}\chi^\beta
-im\frac{p_{\alpha\beta}}{\sqrt{-p^2}}{\rm d}\chi^\alpha\wedge
{\rm d}\chi^\beta\,,  \nonumber
\\ && \label{3.17}
\end{eqnarray}
where $\Pi_{ab}=\eta_{ab}-p_ap_b/p^2$.
It is nondegenerate again if $|b|\neq1$. The superparticle dynamics on
${\cal M}^{6|4}$ is governed by the mass shell constraint $p^2+m^2=0$ only
and provides the straightforward N=2 supergeneralization of the canonical
description of spinning particle. Rel.~(\ref{3.17}) is an N=2 analogue
of the monopole Dirac symplectic structure (\ref{2.1}). It is likely
to be interesting to invert the symplectic two-superform on ${\cal M}^{6|4}$
and to represent the analogue of the fundamental Poisson brackets~(\ref{2.4}).
The nonvanishing brackets are (we mark the PB's on ${\cal M}^{6|4}$ by the
star)
\begin{equation} \begin{array}{c} \displaystyle
\{x^a\,,\,x^b\}^\ast=s\frac{\epsilon^{abc}p_c}{(-p^2)^{3/2}}
\left[1-\frac{m}{2s}
(\theta^\alpha\theta_\alpha+\chi^\alpha\chi_\alpha)+\frac{imb}s\frac{
p_{\alpha\beta}\theta^\alpha\chi^\beta}{\sqrt{-p^2}}\right]\\ \displaystyle
\{\theta^{\alpha\,I}\,,\,\theta^{\beta\,J}\}^\ast=-\frac{1}{2m(1-b^2)}\left(
i\delta^{IJ}\frac{p^{\alpha\beta}}{\sqrt{-p^2}}+b\epsilon^{IJ}
\epsilon^{\alpha\beta}\right)\\
\displaystyle \{x^a\,,\,p_b\}^\ast=\delta^a{}_b\qquad \{x^a\,,\,
\theta^{\alpha\,I}\}^\ast=
-\frac{i}{2p^2}\epsilon^{abc}p_b(\gamma_c)^\alpha{}_\beta\theta^{\beta\,I}
\,.\label{3.18}
\end{array}
\end{equation}
These nonlinear brackets defy the usual attempts of operator realization in a
Hilbert space. An efficient alternative to the direct realization is in the
use of the extended phase superspace
${\cal M}^{8|4}\cong T^\ast({\rm R}^{1,2})\times{\cal L}^{1|2}$, which
allows more supersymmetry, that affects on the quantization procedure
drastically.

%%%%%%%%%%%%%%%%%%%%%%%%%%%%%%%%%%%%%%%%%%%%%%%%%%%%%%%%%%%%%%%%%%%%%%%%
\subsection[Hidden $\rm su(1,1|2)$ supersymmetry of the superspin degrees
of freedom]{\normalsize\bf
Hidden $\bf su(1,1|2)$ supersymmetry of the superspin degrees of
freedom}\hspace*{\parindent}\label{sbs3.4}\nopagebreak
%%%%%%%%%%%%%%%%%%%%%%%%%%%%%%%%%%%%%%%%%%%%%%%%%%%%%%%%%%%%%%%%%%%%%%%%
We have shown that the superparticle dynamics is embedded in the phase
superspace ${\cal M}^{8|4}\cong T^\ast({\rm R}^{1,2})\times{\cal L}^{1|2}$.
One can imply that the inner supermanifold ${\cal L}^{1|2}$ carries internal
(both even and odd) degrees of freedom of D=3 particle. Then the
symplectomorphisms of ${\cal L}^{1|2}$ should be treated as the hidden
supersymmetry of the particle internal structure. Consider this
supersymmetry in more detail. To be specific, let us assume that $|b|<1$. The
degenerate case will be discussed separately in Subsec.~6.

We have already mentioned that ${\cal L}^{1|2}$ is a homogeneous
$\rm OSp(2|2)$ superspace. Introducing new odd complex variables (\ref{3.9})
we established that the symplectic two-superform (\ref{3.10}) reduces
to the superform on the regular $\rm OSp(2|2)$ coadjoint orbit obtained
earlier in Refs.~\cite{Balant,Grad} in the framework of the supercoherent
state technique. A crucial point is that ${\cal L}^{1|2}$ reveals a {\it
K\"ahler\/} supermanifold structure with the superpotential
\begin{equation}
\Phi=-2s\ln(1-z\bar z)-s(1+b)\frac{\theta\bar\theta}{1-z\bar z}
-s(1-b)\frac{\chi\bar\chi}{1-z\bar z}
+\frac{s(1-b^2)}2\frac{\theta\bar\theta\chi\bar\chi}{(1-z\bar z)^2}\,,
\label{3.19}
\end{equation}
so that
\begin{displaymath}
\displaystyle
\Omega_{{\cal L}^{1|2}}=\frac{i}2
\left({\rm d}\bar z\frac\partial{\partial\bar z}
+{\rm d}\bar\theta\frac{\vec\partial}{\partial\bar\theta}+
{\rm d}\bar\chi\frac{\vec\partial}{\partial\bar\chi}\right)\wedge
\left({\rm d}z\frac\partial{\partial z}
+{\rm d}\theta\frac{\vec\partial}{\partial\theta}+
{\rm d}\chi\frac{\vec\partial}{\partial\chi}\right)\Phi\,,
\end{displaymath}
and $\rm OSp(2|2)$ acts on N=2 superunit disc by the {\it superholomorphic}
transformations. Moreover, the supergroup of the superholomorphic
symplectomorphisms of ${\cal L}^{1|2}$ is in fact essentially larger than
$\rm OSp(2|2)$ and it contains at least the supergroup $\rm
SU(1,1|2)$. The corresponding infinitesimal transformations read
\begin{eqnarray}&&\displaystyle \delta
z=i\omega^a\xi_a-\frac{\sqrt{1+b}}2\epsilon_\alpha z^\alpha\theta
-\frac{\sqrt{1-b}}2\eta_\alpha z^\alpha\chi\label{3.20}\\ &&\displaystyle
\delta\theta=\frac{i}2\omega^a\partial\xi_a\theta+\frac{i}{2}
\sqrt\frac{1-b}{1+b}\mu_1\chi-\frac{i}{2}(\mu_2+\mu_3)\theta-
\frac{1}{\sqrt{1+b}}\bar\epsilon_\alpha
z^\alpha-\frac{\sqrt{1-b}}2\eta_\alpha\partial z^\alpha\theta\chi\nonumber\\
&&\displaystyle
\delta\chi=\frac{i}2\omega^a\partial\xi_a\chi-\frac{i}{2}
\sqrt\frac{1+b}{1-b}\bar\mu_1\theta-\frac{i}{2}(\mu_2-\mu_3)\chi
+\frac{\sqrt{1+b}}2\epsilon_\alpha\partial z^\alpha\theta\chi
-\frac{1}{\sqrt{1-b}}\bar\eta_\alpha z^\alpha\,,\nonumber
\end{eqnarray}
where $\partial\equiv\partial/\partial z$, even parameters
$\omega^a,\mu_2,\mu_3$ are real, even parameter $\mu_1$ is complex
and the odd ones $\epsilon_\alpha,\eta_\alpha$ are complex.
Transformations (\ref{3.20}) are generated by the following
Hamiltonians, which may be obtained straightforwardly solving Eqs.\
(\ref{3.11}). There are seven (real) even Hamiltonians
\renewcommand{\theequation}{\arabic{section}.\arabic{equation}a}
\begin{eqnarray}&&\hspace*{-7mm}
\displaystyle
J_a=-sn_a\left(1-\frac{1+b}{2}\frac{\theta\bar\theta}{1-z\bar{z}}
-\frac{1-b}{2}\frac{\chi\bar\chi}{1-z\bar{z}}
+\frac{1-b^2}{2}\frac{\theta\bar\theta\chi\bar\chi}{(1-z\bar{z})^2}
\right)\nonumber\\&&\hspace*{-7mm}
\displaystyle
P_1=s\frac{\sqrt{1-b^2}}{2}\frac{\theta\bar\chi-\bar\theta\chi}{1-z\bar{z}}
\quad\hspace{2mm} P_3=-s\left(\frac{1+b}{2}\frac{\theta\bar\theta}{1-z\bar{z}}
-\frac{1-b}{2}\frac{\chi\bar\chi}{1-z\bar{z}}\right)\label{3.21a}\\&&
\displaystyle\hspace*{-7mm}
P_2=is\frac{\sqrt{1-b^2}}{2}\frac{\theta\bar\chi+\bar\theta\chi}{1-z\bar{z}}
\quad P_4=-s\left(\frac{1+b}{2}\frac{\theta\bar\theta}{1-z\bar{z}}
+\frac{1-b}{2}\frac{\chi\bar\chi}{1-z\bar{z}}
-\frac{1-b^2}{2}\frac{\theta\bar\theta\chi\bar\chi}{(1-z\bar{z})^2}\right)
\nonumber
\end{eqnarray}\addtocounter{equation}{-1}%
\renewcommand{\theequation}{\arabic{section}.\arabic{equation}b}%
and eight odd ones
\begin{equation}
\begin{array}{ll}
\displaystyle
E^\alpha=s\sqrt{1+b}
\left(\frac{z^\alpha\bar\theta-\bar z^\alpha\theta}{1-z\bar z}\right)
\left(1-\frac{1-b}{2}
\frac{\chi\bar\chi}{1-z\bar{z}}\right) & \displaystyle
F^\alpha=in^\alpha{}_\beta E^\beta \\
\displaystyle
G^\alpha=s\sqrt{1-b}
\left(\frac{z^\alpha\bar\chi-\bar z^\alpha\chi}{1-z\bar z}\right)
\left(1-\frac{1+b}{2}
\frac{\theta\bar\theta}{1-z\bar{z}}\right) & \displaystyle
H^\alpha=in^\alpha{}_\beta G^\beta\,.
\end{array}
\label{3.21b}
\end{equation}\renewcommand{\theequation}{\arabic{section}.\arabic{equation}}%
\sbox{\innhamnuber}{(\theequation)}%
These Hamiltonians, together with one more even element $Z\equiv s$, generate
a closed superalgebra with respect to Poisson superbrackets on
${\cal L}^{1|2}$ (here $I,J,K=1,2,3$):
\begin{eqnarray}&&\displaystyle\hspace*{-5mm}
\begin{array}{lll}
\{J_a,J_b\}=\epsilon_{abc}J^c &
\displaystyle \{J_a,E^\alpha\}=\frac{i}{2}(\gamma_a)^\alpha{}_\beta E^\beta &
\displaystyle \{J_a,F^\alpha\}=\frac{i}{2}(\gamma_a)^\alpha{}_\beta F^\beta
\\[2mm]
\{P_I,P_J\}=-\epsilon_{IJK}P_K &
\displaystyle \{J_a,G^\alpha\}=\frac{i}{2}(\gamma_a)^\alpha{}_\beta G^\beta &
\displaystyle \{J_a,H^\alpha\}=\frac{i}{2}(\gamma_a)^\alpha{}_\beta H^\beta
\end{array}
\nonumber\\&&\hspace*{-5mm}
\begin{array}{llll}
\displaystyle \{E^\alpha,P_1\}=\frac12 H^\alpha&
\displaystyle \{E^\alpha,P_2\}=-\frac12 G^\alpha&
\displaystyle \{E^\alpha,P_3\}=-\frac12 F^\alpha&
\displaystyle \{E^\alpha,P_4\}=-\frac12 F^\alpha\\[2mm]
\displaystyle \{F^\alpha,P_1\}=-\frac12 G^\alpha&
\displaystyle \{F^\alpha,P_2\}=-\frac12 H^\alpha&
\displaystyle \{F^\alpha,P_3\}=\frac12 E^\alpha&
\displaystyle \{F^\alpha,P_4\}=\frac12 E^\alpha\\[2mm]
\displaystyle \{G^\alpha,P_1\}=\frac12 F^\alpha&
\displaystyle \{G^\alpha,P_2\}=\frac12 E^\alpha&
\displaystyle \{G^\alpha,P_3\}=\frac12 H^\alpha&
\displaystyle \{G^\alpha,P_4\}=-\frac12 H^\alpha\\[2mm]
\displaystyle \{H^\alpha,P_1\}=-\frac12 E^\alpha&
\displaystyle \{H^\alpha,P_2\}=\frac12 F^\alpha&
\displaystyle \{H^\alpha,P_3\}=-\frac12 G^\alpha&
\displaystyle \{H^\alpha,P_4\}=\frac12 G^\alpha\\[3mm]
\end{array}
\nonumber\\&&\displaystyle\hspace*{-5mm}
\begin{array}{lll}
\displaystyle \{E^\alpha,F^\beta\}=\epsilon^{\alpha\beta}(Z-P_3) &
\displaystyle \{E^\alpha,G^\beta\}=-\epsilon^{\alpha\beta}P_2 &
\displaystyle \{E^\alpha,H^\beta\}=\epsilon^{\alpha\beta}P_1 \\[2mm]
\displaystyle \{G^\alpha,H^\beta\}=\epsilon^{\alpha\beta}(Z+P_3) &
\displaystyle \{F^\alpha,H^\beta\}=-\epsilon^{\alpha\beta}P_2 &
\displaystyle \{F^\alpha,G^\beta\}=\epsilon^{\alpha\beta}P_1\\[1.5mm]
\end{array}\label{3.22}
\\&&\hspace*{-5mm} \displaystyle \ \,
\{E^\alpha,E^\beta\}=\{F^\alpha,F^\beta\}=
\{G^\alpha,G^\beta\}=\{H^\alpha,H^\beta\}=i(\gamma_a)^{\alpha\beta}J^a
\nonumber \\[1.5mm]&&\hspace*{-5mm}\displaystyle\ \,
\{J_a,P_I\}=0\quad \{P_I,P_4\}=0\quad \{J_a,P_4\}=0\quad
\{Z,{\rm anything}\}=0\,.\nonumber
\end{eqnarray}
What we have obtained it is the explicit Poisson realization of the so-called
$\rm su(1,1|2)$ superalgebra~\cite{Dictionary}, whose even part is $\rm
su(1,1|2)_0=su(1,1) \bigoplus u(2)\bigoplus R$ and the odd part constitutes
an eight dimensional module of the even part; $Z$ presents a central charge.
The $\rm osp(2|2)$ subsuperalgebra found in Refs.~\cite{Balant,Grad}
is spanned by $J_a,B,\sqrt{ms}\,V^\alpha,\sqrt{ms}\,W^\alpha$, where
$V^\alpha,W^\alpha$ are defined below by Eqs.\ (\ref{3.24}) and $B=P_3-bZ$.
We reveal that $\rm N=2$ superunit disc is not only a typical
coadjoint orbit of the $\rm OSp(2|2)$ supergroup, ${\cal
L}^{1|2}\cong \rm OSp(2|2)/[U(1)\times U(1)]$, but it can be
treated  simultaneously as an atypical K\"ahler orbit of the supergroup $\rm SU(1,1|2)$:
${\cal L}^{1|2}\cong \rm SU(1,1|2)/[U(2|2)\times U(1)]$.

%%%%%%%%%%%%%%%%%%%%%%%%%%%%%%%%%%%%%%%%%%%%%%%%%%%%%%%%%%%%%%%%%%%%%%%%%
\subsection[Hidden N=4 Poincar\'e supersymmetry]
{\normalsize\bf\hspace{-2mm}Hidden N=4 Poincar\'e supersymmetry}
\hspace*{\parindent}\label{sbs3.5}\nopagebreak%%%%%%%%%%%%%%%%%%%%%%%%%%%
%%%%%%%%%%%%%%%%%%%%%%%%%%%%%%%%%%%%%%%%%%%%%%%%%%%%%%%%%%%%%%%%%%%%%%%%%
Subalgebra $\rm u(2)$ of the internal $\rm su(1,1|2)$ superalgebra
acts on the odd variables, as is seen from Eqs.\
(\ref{3.20}). It is exactly the subalgebra of the isotopic symmetry.
However, the isotopic~$\rm U(2)$ symmetry may now be involved in the
Poincar\'e supersymmetry.  The isotopic rotations together with the N=2
Poincar\'e transformations~(\ref{3.2}),~(\ref{3.3}) generate (when
$|b|\neq1$) more wide D=3, N=4 Poincar\'e superalgebra. In addition to
(\ref{3.2}) there are the following supersymmetry transformations:
\begin{equation}
\begin{array}{lll}
\displaystyle\hspace*{-1.45mm}
\delta_{\tilde\epsilon}x^a=
ib\gamma^a_{\alpha\beta}\tilde\epsilon^\alpha\chi^\beta
-i\epsilon^{abc} n_b \gamma_{c\,\alpha\beta}\tilde\epsilon^\alpha\theta^\beta+
n^a\tilde\epsilon^\alpha\theta_\alpha &
\displaystyle
\delta_{\tilde\epsilon}\theta^\alpha=-in^\alpha{}_\beta\tilde\epsilon^\beta &
\displaystyle
\delta_{\tilde\epsilon} \chi^\alpha=
\delta_{\tilde\epsilon} z=0 \\
\displaystyle \hspace*{-1.45mm}
\delta_{\tilde\eta}x^a=-ib\gamma^a_{\alpha\beta}\tilde\eta^\alpha\theta^\beta
-i\epsilon^{abc}n_b \gamma_{c\,\alpha\beta}\tilde\eta^\alpha\chi^\beta+
n^a\tilde\eta^\alpha\chi_\alpha &
\displaystyle
\delta_{\tilde\eta}\chi^\alpha=-in^\alpha{}_\beta\tilde\eta^\beta &
\displaystyle
\delta_{\tilde\eta}\theta^\alpha=
\delta_{\tilde\eta} z=0\,,
\end{array}
\label{3.2a}
\end{equation}
where $\tilde\epsilon^\alpha,\tilde\eta^\alpha$ are odd infinitesimal
parameters. The respective Hamiltonians on ${\cal M}^{8|4}$ read
\begin{equation}
\begin{array}{l}\displaystyle
\tilde{\cal Q}^1_\alpha=
ip_{\alpha\beta}(in^\beta{}_\gamma\theta^\gamma+b\chi^\beta)
-m(\theta_\alpha-ibn_{\alpha\beta}\chi^\beta)\\ \displaystyle
\tilde{\cal Q}^2_\alpha=
ip_{\alpha\beta}(in^\beta{}_\gamma\chi^\gamma-b\theta^\beta)
-m(\chi_\alpha+ibn_{\alpha\beta}\theta^\beta)\,.
\end{array}
\label{3.12a}
\end{equation}

New supercharges together with N=2 superpoincar\'e Hamiltonians~(\ref{3.12})
and isotopic~$\rm U(2)$ Hamiltonians $P_I\,,I=1,2,3,4$ generate closed N=4
Poincar\'e superalgebra with one central charge. It can be seen by introducing
new basis for supercharges~$2{\cal R}^I_\alpha=({\cal Q}^1_\alpha+\tilde{\cal
 Q}^2_\alpha\,,\,{\cal Q}^2_\alpha-\tilde{\cal Q}^1_\alpha)$,
$2\tilde{\cal R}^I_\alpha=(\tilde{\cal Q}^1_\alpha+{\cal Q}^2_\alpha\,,\,
\tilde{\cal Q}^2_\alpha-{\cal Q}^1_\alpha)\,, I=1,2$. On shell~(\ref{3.6})
we have
\begin{equation}
\begin{array}{l}\displaystyle
\{{\cal R}^I_\alpha\;,\;{\cal R}^J_\beta\}\approx(1-b)(
-i\delta^{IJ}p_{\alpha\beta}+m\epsilon^{IJ}\epsilon_{\alpha\beta})\\
\displaystyle\{\tilde{\cal R}^I_\alpha\;,\;\tilde{\cal
R}^J_\beta\}\approx(1+b)
(-i\delta^{IJ}p_{\alpha\beta}+m\epsilon^{IJ}\epsilon_{\alpha\beta})\\
\displaystyle\{{\cal R}^I_\alpha\;,\;\tilde{\cal R}^J_\beta\}\approx0\,.
\end{array}
\end{equation}

The invariance of the original Lagrangian
(\ref{3.1}) under the transformations (\ref{3.2a}) can be examined
straightforwardly. Thus, the model, being N=2 superpoincar\'e invariant by
construction, allows the hidden N=4 supersymmetry. The appearance the enhanced
supersymmetry is hardly surprising  in the model. This
N=4 supersymmetry is degenerate in a sense
that the corresponding central charges
equals to $m$ and, so, they saturate the BPS bound for N=4 Poincar\'e
superalgebra. It reflects the degeneracy of N=4 supersymmetry and the
shortening of the N=4 superparticle multiplet to the N=2 supermultiplet in
quantum theory. Moreover, it is a general property of extended supersymmetry
that some of the degenerate multiplets of a larger SUSY (those which
saturate the BPS bound) have the same particle content, as is observed
in the respective
multiplets of a  smaller SUSY. This fact provides a simple reason why
some of supersymmetric theories may  have  the extended supersymmetries.
The precedents
are known both for D=4,6,10 superparticle models~\cite{Brink} and
supersymmetric field theories (for example, the theories with non-trivial
topological charge~\cite{SpecHlou}). D=3, N=1 superparticle allows the hidden
N=2 SUSY~\cite{GKL2}.

The degeneracy of the hidden N=4 supersymmetry can be observed already in the
classical model. New supercharges $\tilde{\cal Q}^I_\alpha$ are functionally
dependent from the N=2 Hamiltonians. On the constraint surface (\ref{3.6}) we
have
\begin{equation}
\tilde{\cal Q}^I_\alpha=-\frac{i}{m}p_\alpha{}^\beta{\cal Q}^I_\beta\,.
\label{relN4}
\end{equation}
We conclude that the hidden N=4 supersymmetry can be treated as an artifact
of the embedding of N=4 Poincar\'e superalgebra into the universal enveloping
algebra of N=2 one. The transformations (\ref{3.2a}) are, in fact,
special linear combinations of the N=2 transformations (\ref{3.2}) with the
coefficients depending from the on shell conserved quantities.

%%%%%%%%%%%%%%%%%%%%%%%%%%%%%%%%%%%%%%%%%%%%%%%%%%%%%%%%%%%%%%%%%%%%%%%%
\subsection[Bogomol'ny-Prassad-Sommerfield limit]
{\normalsize\bf\hspace{-4mm}
Bogomol'ny-Prassad-Sommerfield limit\hspace*{-2mm}}
\hspace*{\parindent}\label{sbs3.6}\nopagebreak
%%%%%%%%%%%%%%%%%%%%%%%%%%%%%%%%%%%%%%%%%%%%%%%%%%%%%%%%%%%%%%%%%%%%%%%%%
Let us  briefly discuss the special case $|b|=1$. To make the
mentioned degeneracy more evident we introduce for a while new odd variables
\begin{displaymath}
\tilde\theta^\alpha=\theta^\alpha-in^\alpha{}_\beta\chi^\beta\qquad
\tilde\chi^\alpha=\chi^\alpha-in^\alpha{}_\beta\theta^\beta
\end{displaymath}
instead of $\theta^\alpha,\chi^\alpha$. This change of the odd variables
is one-to-one, and the original Lagrangian (\ref{3.1}) reads in new variables
as
$$
L=m(\dot{x},n)-im\frac{1+b}{2}n_{\alpha\beta}\tilde\theta^\alpha
\dot{\tilde\theta}{}^\beta-im\frac{1-b}{2}n_{\alpha\beta}\tilde\chi^\alpha
\dot{\tilde\chi}{}^\beta+is\frac{\bar z{\dot z} -z\dot{\bar z}}{1-z\bar z}\,.
\eqno{(\rm 3.1a)}
$$
It is seen immediately that half of the odd degrees of freedom of the
superparticle drops out from the theory in the case of $|b|=1$. Moreover, in
the BPS limit expression~(3.1a) reduces to the Lagrangian of N=1, D=3
superparticle \cite{GKL2} and does describe not a superquartet, but a
supersymmetric doublet of particles of equal mass $m$ and spins $s$ and
$s+\frac12$ only.

The inner symplectic two-superform (\ref{3.10}) turns out to be
a K\"ahler superform on an atypical $\rm OSp(2|2)$-coadjoint orbit ${\cal
L}^{1|1}$ of complex dimension $1/1$. Thus the phase superspace
${\cal M}^{8|4}$ reduces to ${\cal M}^{8|2}\cong T^\ast({\rm R}^{1,2})
\times{\cal L}^{1|1}$ with internal $\rm OSp(2|2)$ supersymmetry. The
latter is realized by all superholomorphic transformations of~N=1 superunit
disc ${\cal L}^{1|1}$. The respective model of N=1, D=3 superparticle on
${\cal M}^{8|2}$ was considered in detail in Ref.~\cite{GKL2}. In the
present paper we give an appropriate~N=2 extension for the N=1 model
retaining the hidden supersymmetries.

It is worth noting that the hidden N=4 supersymmetry vanishes if $|b|=1$,
whereas~N=2 supersymmetry of the N=1 superparticle could be treated as the
hidden one \cite{GKL2}.  Almost all the equations of this Section still
remain valid in the BPS limit if one takes formally $\tilde\theta^\alpha=
\theta^\alpha\,,\chi^\alpha=0$ and $b=0$.

%%%%%%%%%%%%%%%%%%%%%%%%%%%%%%%%%%%%%%%%%%%%%%%%%%%%%%%%%%%%%%%%%%%%%%%%
\subsection[Relationship between Hamiltonian generators of the Poincar\'e and
internal supersymmetries]
{\normalsize\bf\hspace{-4mm} Relationship between
Hamiltonian generators of the Poincar\'e \protect\\ \hspace*{-2mm}and
internal supersymmetries} \hspace*{\parindent}\label{sbs3.7}\nopagebreak
%%%%%%%%%%%%%%%%%%%%%%%%%%%%%%%%%%%%%%%%%%%%%%%%%%%%%%%%%%%%%%%%%%%%%%%%%
We have observed that the model contains both the global Poincar\'e SUSY and
the hidden $\rm SU(1,1|2)$, the latter is closely related to the superspin
intrinsic structure. Thus, the relevant quantization
procedure should make a provision for either symmetries to survive
in quantum theory. This quantization can be based on a simple fact that the
Hamiltonian generators (\ref{3.12}) and (\ref{3.12a}) of the Poincar\'e
supersymmetries, being the functions on ${\cal M}^{8|4}\cong T^\ast({\rm
R}^{1,2})\times {\cal L}^{1|2}$, can be expressed in terms of the
Minkowski-space co-ordinates and momenta $(x^a,p_a)$ and of the $\rm
su(1,1|2)$ Hamiltonians $J_a,P_I,E^\alpha,F^\alpha,G^\alpha$ and $H^\alpha$
\usebox{\innhamnuber}, which parametrize the coadjoint orbit ${\cal
L}^{1|2}$. We give here the explicit form of these expressions:
\begin{equation}
\begin{array}{l}\displaystyle
{\cal J}_a=\epsilon_{abc}x^bp^c+J_a\qquad\qquad {\cal P}_a=p_a\qquad\qquad
{\cal Z}=mb\\[1mm]
\displaystyle
{\cal Q}^1_\alpha=(ip_{\alpha\beta}W^\beta+m\tilde{W}_\alpha)
[1+{\rm q}^{cl}(bP_3-\sqrt{1-b^2}\,P_2-P_4)] \\[1mm] \displaystyle
{\cal Q}^2_\alpha=(ip_{\alpha\beta}V^\beta+m\tilde{V}_\alpha)
[1+{\rm q}^{cl}(bP_3+\sqrt{1-b^2}\,P_2-P_4)]
\end{array}
\label{3.23}
\end{equation}
\vspace*{-2mm}
\begin{equation}
\begin{array}{l}\displaystyle
\displaystyle
\tilde{\cal Q}^1_\alpha=(ip_{\alpha\beta}\tilde W^\beta-mW_\alpha)
[1+{\rm q}^{cl}(bP_3-\sqrt{1-b^2}\,P_2-P_4)] \\[1mm] \displaystyle
\tilde{\cal Q}^2_\alpha=(ip_{\alpha\beta}\tilde V^\beta-mV_\alpha)
[1+{\rm q}^{cl}(bP_3+\sqrt{1-b^2}\,P_2-P_4)]\,.
\end{array}
\label{N4}
\end{equation}
where
\begin{equation}
\begin{array}{ll}
\displaystyle
W^\alpha=\frac{1}{2\sqrt{ms}}(\sqrt{1+b}\,E^\alpha+\sqrt{1-b}\,H^\alpha) &
\displaystyle\tilde W^\alpha=
\frac{1}{2\sqrt{ms}}(\sqrt{1+b}\,F^\alpha-\sqrt{1-b}\,G^\alpha)\\
\displaystyle
V^\alpha=\frac{1}{2\sqrt{ms}}(\sqrt{1+b}\,F^\alpha+\sqrt{1-b}\,G^\alpha) &
\displaystyle\tilde V^\alpha=
\frac{1}{2\sqrt{ms}}(\sqrt{1-b}\,H^\alpha-\sqrt{1+b}\,E^\alpha)
\end{array}
\label{3.24}
\end{equation}
and constant ${\rm q}^{cl}$ reads as
\begin{equation}
{\rm q}^{cl}=\frac{1}{4s}\label{3.25}\,.
\end{equation}
To construct an appropriate operator realization of these expressions
we shall quantize $(x^a,p_a)$ canonically and extend simultaneously $\rm
su(1,1|2)$ Hamiltonian vector fields to a representation by Hermitian
operators in Hilbert space.

Notice at once some important details in relation to the quantization
which should be compatible to the full symmetry of the superparticle.
First, expressions~(\ref{3.23}) and~(\ref{N4}) are essentially {\it
nonlinear} in the generators \usebox{\innhamnuber} of the inner $\rm
su(1,1|2)$ superalgebra. Thus, even though the operator realization of the
Poisson $\rm su(1,1|2)$ superalgebra~(\ref{3.22}) is found and the
corresponding operators are substituted in Eq.~(\ref{3.23}), we may not be
sure that the representation of the Poincar\'e superalgebra (neither N=2
nor N=4) is reproduced for certain in quantum theory. Because of the
nonlinearity, the superalgebra of operators, corresponding to
(\ref{3.23}), might be disclosed, and it is the parameter~${\rm q}$, which
controls the possible disclosure of the Poincar\'e superalgebra. We will
see that the parameter~${\rm q}^{cl}$ should be renormalized in quantum
theory to reproduce a representation of the Poincar\'e supersymmetry.

Second, it is a matter of direct verification that the Hamiltonians
$W^\alpha,\tilde W^\alpha$ have vanishing Poisson superbrackets with
$bP_3-\sqrt{1-b^2}P_2-P_4$, whereas $V^\alpha,\tilde V^\alpha$ commute to
$bP_3+\sqrt{1-b^2}\,P_2-P_4$. This point will be important for
Hermitian properties of operators in quantum mechanics.

\section[\hspace*{2mm} First quantization of the superparticle]
{\normalsize\bf\hspace{-5mm}FIRST QUANTIZATION OF THE
SUPERPARTICLE}\label{s2}\hspace*{\parindent}\setcounter{equation}{0}%
%%%%%%%%%%%%%%%%%%%%%%%%%%%%%%%%%%%%%%%%%%%%%%%%%%%%%%%%%%%%%%%%%%%%%%%%%%
It is a primary objective of previous consideration to present the classical
model of N=2, D=3 superparticle in the form, well-adapted for a
quantizing procedure. We have obtained an embedding of the (maximal)
coadjoint orbit of N=2 Poincar\'e supergroup in the extended phase
superspace ${\cal M}^{8|4}\cong T^\ast({\rm R}^{1,2})\times {\cal
L}^{1|2}$. Going to quantum theory we will combine the canonical Dirac
quantization on $T^\ast({\rm R}^{1,2})$ and the geometric quantization
methods on the $\rm SU(1,1|2)$ co-orbit ${\cal L}^{1|2}$. In particular, a
combination of the standard real polarization in $T^\ast({\rm R}^{1,2})$
and the K\"ahler one in ${\cal L}^{1|2}$ will be used to construct the
superparticle's Hilbert space.

The quantization scheme implies from the outset that the internal $\rm
SU(1,1|2)$ supersymmetry must survive at the quantum level. Mutual relation
between Hamiltonians of $\rm SU(1,1|2)$ and Poincar\'e supersymmetries, being
expressed by Eq.~(\ref{3.23}) and~(\ref{N4}), is crucial in our approach.
At first, we construct the operator realization for the Hamiltonians of
$\rm su(1,1|2)$ superalgebra in the framework of Berezin quantization.
Then the expressions~(\ref{3.23}) (possibly, together with~(\ref{N4})) are
used to obtain the realization of a UIR for the N=2 (respectively,
enhanced N=4) Poincar\'e superalgebra.  We find that the classical meaning
of the parameter $\rm q$ (\ref{3.25}) in the relations~(\ref{3.23})
and~(\ref{N4}) should be accompanied by certain quantum corrections,
referred to as a renormalization, for consistency of the quantum theory.

Eventually we obtain the straightforward N=2 supergeneralization
of conventional realization of the unitary irreducible representations
(UIR's) of D=3 Poincar\'e group on the fields carrying representations of
$\rm\overline{SO^\uparrow(1,2)}$~\cite{JackNair}. Two cases should
be distinguished among these representations. The fields describing
fractional superspin (superanyons) carry an atypical unitary {\it infinite
dimensional\/} UIR's of $\rm SU(1,1|2)$, whereas the UIR's of (half)integer
superspin can be realized on the spin-tensor fields carrying atypical
{\it finite dimensional\/} non-unitary representations of $\rm SU(1,1|2)$.
The realization of the superparticle Hilbert space is slightly different
in these two cases.

%%%%%%%%%%%%%%%%%%%%%%%%%%%%%%%%%%%%%%%%%%%%%%%%%%%%%%%%%%%%%%%%%%%%%%%%
\subsection[Berezin quantization on ${\cal L}^{1|2}$.]
{\normalsize\bf\hspace{-2mm}Berezin quantization on ${\cal L}^{1|2}$.
\protect\\ \hspace*{-2mm}}\label{sbs4.1}
\hspace*{\parindent}\nopagebreak
%%%%%%%%%%%%%%%%%%%%%%%%%%%%%%%%%%%%%%%%%%%%%%%%%%%%%%%%%%%%%%%%%%%%%%%%%
The Berezin technique~\cite{Berezin1,Berezin2,Perel,BarMar} provides the
perfect quantization method for the K\"ahler homogeneous spaces. We consider
here briefly the application of this method to the supermanifold $\rm{\cal
L}^{1|2}\cong OSp(2|2)/[U(1)\times U(1)]\cong SU(1,1|2)/[U(2|2)\times U(1)]$
with the nondegenerate symplectic structure when $|b|<1$. The geometric
quantization on ${\cal L}^{1|2}$, being considered as a regular coadjoint
orbit, is studied in~Refs.~\cite{Balant,Grad} in detail. However, as we
know, ${\cal L}^{1|2}$ has not  been considered as an irregular
$\rm SU(1,1|2)$ co-orbit nor as a detailed Berezin quantization and the
underlying correspondence principle is not explicitly established.

In the following Subsections we apply the obtained results
for quantization of D=3 superparticle.

\subsubsection{Antiholomorphic sections and an inner
product} \hspace*{\parindent}\nopagebreak
%%%%%%%%%%%%%%%%%%%%%%%%%%%%%%%%%%%%%%%%%%%%%%%%%%%%%%%%%%%%%%%%%%%%%%%%%%
Let us consider the space ${\cal O}_{s,b}$ of
superantiholomorphic sections of the superholomorphic line bundle over
${\cal L}^{1|2}$, whose elements are represented by functions
\begin{eqnarray}\displaystyle
f(\bar{\mit\Gamma})\equiv
f(\bar{z},\bar{\theta},\bar{\chi})&\displaystyle =& \displaystyle
f_0(\bar{z})+\sqrt{s(1+b)}\,\bar\theta f_1(\bar{z}) +\sqrt{s(1-b)}\,\bar\chi
f_2(\bar{z})\label{4.1}\\&&\displaystyle
+\sqrt{s(s+1/2)(1-b^2)}\,\bar\theta\bar\chi f_3(\bar{z})\,,\nonumber
\end{eqnarray}
where $f_i(\bar z)\,,i=0,1,2,3$ are ordinary antiholomorphic functions on the
unit disc of  the complex plane. We denote by
${\mit\Gamma}\equiv\{{\mit\Gamma}^A\}=\{z,\theta,\chi\}$ and
$\bar{\mit\Gamma}\equiv\{{\mit\Gamma}^{\bar A}\}=\{\bar
z,\bar\theta,\bar\chi\}$ the sets of the superholomorphic and
superantiholomorphic variables respectively. The space ${\cal O}_{s,b}$ is
equipped naturally by an inner product
\renewcommand{\theequation}{\arabic{section}.\arabic{equation}a}
\begin{equation}
\langle f|g\rangle_{{\cal L}^{1|2}}=\int_{{\cal
L}^{1|2}}\overline{f(\bar{\mit\Gamma})}g(\bar{\mit\Gamma})
{\rm e}^{-\Phi({\mit\Gamma},\bar{\mit\Gamma})}
d\mu({\mit\Gamma},\bar{\mit\Gamma})\,.\label{4.2a}
\end{equation}\renewcommand{\theequation}{\arabic{section}.\arabic{equation}}%
Here $\Phi({\mit\Gamma},\bar{\mit\Gamma})$ is the K\"ahler superpotential
(\ref{3.19}) and $d\mu({\mit\Gamma},\bar{\mit\Gamma})$ is an $\rm SU(1,1|2)$
invariant Liouville supermeasure on ${\cal L}^{1|2}$. Taking into account
the definition of the symplectic two-superform (\ref{3.10})
$\Omega_{{\cal L}^{1|2}}\equiv{\rm d}{\mit\Gamma}^A\Omega_{A\bar{B}}{\rm d}
{\mit\Gamma}^{\bar B}$, one can derive the supermeasure
explicitly~\cite{Balant,Grad}
\begin{equation}
\hspace*{-2.5mm}
d\mu({\mit\Gamma},\bar{\mit\Gamma})=-\frac{1}{4\pi}{\rm sdet}
\|\Omega_{A\bar{B}}\|d{\mit\Gamma}\, d\bar{\mit\Gamma}
=\frac{d{\mit\Gamma}\,d\bar{\mit\Gamma}}{i\pi s(1-b^2)} \, ,
\qquad\!
d{\mit\Gamma}\, d\bar{\mit\Gamma}\equiv dz\,d\bar z\,d\theta\,
d\bar\theta\,d\chi\,d\bar\chi\,.\!\!
\label{4.3}
\end{equation}
Using Eqs.~(\ref{3.19}), (\ref{4.1}) and (\ref{4.3}) we can integrate out
the Grassmann variables in Eq.~(\ref{4.2a}), that
reduces the inner product to the following form:
\addtocounter{equation}{-2}
\renewcommand{\theequation}{\arabic{section}.\arabic{equation}b}
\begin{equation}
\langle f|g\rangle_{{\cal L}^{1|2}}=\langle f_0|g_0\rangle_{\cal L}^s
+\langle f_1|g_1\rangle_{\cal L}^{s+1/2}
+\langle f_2|g_2\rangle_{\cal L}^{s+1/2}+\langle f_3|g_3\rangle_{\cal L}^{s+1}
\,,\label{4.2b}
\end{equation}
\renewcommand{\theequation}{\arabic{section}.\arabic{equation}}%
\sbox{\scalprod}{(\theequation)}\addtocounter{equation}{1}%
where
\begin{equation}
\langle\varphi |\chi\rangle_{{\cal L}}^l=(2l-1)
\int\limits_{|z|<1}\frac{dzd\bar{z}}{2\pi i}(1-z\bar z)^{2l-2}
\overline{\varphi (\bar{z})}\chi (\bar{z}) \label{4.4}
\end{equation}
is an inner product in the representation space $D^l_+$ of
$\rm\overline{SO^\uparrow(1,2)}$ discrete series
bounded below, being realized by
antiholomorphic functions\footnote{The monomials $\phi_n^l=[\Gamma(2l+n)/
\Gamma(n-1)\Gamma(2l)]^{1/2}\bar z^n$, $n$ is integer  non-negative, serve as a
standard orthonormal basis in $D^l_+$.} in the unit disc $|z|<1$. The inner
product~(4.4) is well defined and positive if $l>1/2$. Moreover, for
values $0<l<1/2$ one can still use Eq.\ (\ref{4.4}) if suitable analytic
continuations are made. The case of~$l=1/2$ should be understood in the
sense of the limit. We conclude that the inner product \usebox{\scalprod} in
${\cal O}_{s,b}$ is well defined if $s>0$ (and, of course, if $|b|<1$).

In view of the transformation law for K\"ahler superpotential
$\Phi({\mit\Gamma},\bar{\mit\Gamma})$ under the action of $\rm SU(1,1|2)$
supergroup, the inner product \usebox{\scalprod} holds to be $\rm SU(1,1|2)$
invariant, if an appropriate transformation law for $f({\mit\Gamma})\in{\cal
O}_{s,b}$ is implemented. In other terms, the Hamiltonian action of $\rm
SU(1,1|2)$ on ${\cal L}^{1|2}$ can be lifted to a unitary representation
in~${\cal O}_{s,b}$. We give below an infinitesimal form of this
representation only, that is explicit representation of corresponding
superalgebra $\rm su(1,1|2)$.  To obtain it, we first consider a
conventional correspondence between linear operators in ${\cal O}_{s,b}$ and
Berezin's symbols.

\subsubsection{Classical observables and operators}
\hspace*{\parindent}\nopagebreak
%%%%%%%%%%%%%%%%%%%%%%%%%%%%%%%%%%%%%%%%%%%%%%%%%%%%%%%%%%%%%%%%%%%%%%%%%%%%
Let $A({\mit\Gamma},\bar{\mit\Gamma})$ be a ``classical observable'', that
means it is a real function on ${\cal L}^{1|2}$ to be continuously
differentiable in $z,\bar z$ that the integrals considered below do
exist. We associate a linear operator $\widehat A$ in ${\cal O}_{s,b}$ to the
classical observable $A({\mit\Gamma},\bar{\mit\Gamma})$ by the rule
\begin{equation}
({\widehat A}f)(\bar{\mit\Gamma})=\int_{{\cal L}^{1|2}}
A({\mit\Gamma_1},\bar{\mit\Gamma})f(\bar{\mit\Gamma}_1)
L_{s,b}({\mit\Gamma}_1,\bar{\mit\Gamma})
{\rm e}^{-\Phi({\mit\Gamma}_1,\bar{\mit\Gamma}_1)}
d\mu({\mit\Gamma}_1,\bar{\mit\Gamma}_1)\,.\label{4.5}
\end{equation}
where $A({\mit\Gamma_1},\bar{\mit\Gamma})$ serves only as an analytic
continuation in ${\cal L}^{1|2}\times{\cal L}^{1|2}$ for classical observable
$A({\mit\Gamma},\bar{\mit\Gamma})$. The generating kernel
$L_{s,b}({\mit\Gamma}_1,\bar{\mit\Gamma})$ can be constructed by the use of
an arbitrary complete orthonormal basis $f_k(\bar{\mit\Gamma})$ in ${\cal
O}_{s,b}$, and appears to be related immediately to the analytic continuation
in ${\cal L}^{1|2}\times{\cal L}^{1|2}$ of the K\"ahler superpotential:
\begin{eqnarray}
&\displaystyle\hspace*{-4mm}
L_{s,b}({\mit\Gamma}_1,\bar{\mit\Gamma})& \displaystyle
=\sum\limits_{k=1}^\infty
f_k(\bar{\mit\Gamma})\overline{f_k(\bar{\mit\Gamma}_1)}
=(1-z_{\scriptscriptstyle 1}\bar
z)^{-2s}\left[1-s(1+b)\frac{\theta_{\scriptscriptstyle
1}\bar\theta}{1-z_{\scriptscriptstyle 1}\bar z}
-s(1-b)\frac{\chi_{\scriptscriptstyle 1}\bar\chi}{1-z_{\scriptscriptstyle 1}
\bar z}\right.\nonumber \\ &&\displaystyle\left.
+s(s+\frac12)(1-b^2)
\frac{\theta_{\scriptscriptstyle 1}\bar\theta\chi_{\scriptscriptstyle 1}
\bar\chi}{(1-z_{\scriptscriptstyle 1}\bar z)^2}\right]
=\exp[\Phi({\mit\Gamma}_1,\bar{\mit\Gamma})]\,. \label{4.6}
\end{eqnarray}
The state, being presented by the function $\Phi_{\bar{\mit\Gamma}}
(\bar{\mit\Gamma}_1)=L_{s,b}({\mit\Gamma},\bar{\mit\Gamma}_1)$ with fixed
$\bar{\mit\Gamma}\equiv\{\bar z,\bar\theta,\bar\chi\}$ is denoted by
$|\bar z,\bar\theta,\bar\chi\rangle$, is called as an $\rm SU(1,1|2)$
(or $\rm OSp(2|2)$) supercoherent state. The analytic continuation
in ${\cal L}^{1|2}\times{\cal L}^{1|2}$ for any classical observable could be
expressed in terms of the supercoherent states as follows
\begin{equation}
A({\mit\Gamma_1},\bar{\mit\Gamma_2})=\frac{\langle\Phi_{\bar{\mit\Gamma}_2}
|{\widehat A}|\Phi_{\bar{\mit\Gamma}_1}\rangle
\lefteqn{{}_{{\cal L}^{1|2}}}}
{\langle\Phi_{\bar{\mit\Gamma}_2}|\Phi_{\bar{\mit\Gamma}_1}\rangle
\lefteqn{{}_{{\cal L}^{1|2}}}}
\qquad.\label{4.7}
\end{equation}
So, the symbol of the unit operator $\hat{\rm I}$ is just $1$.
Hence, the one-to-one correspondence between classical observables on
${\cal L}^{1|2}$ and linear operators in ${\cal O}_{s,b}$ is established.
In view of Eq.\ (\ref{4.7}) classical observables are also referred to
as (covariant) Berezin symbols.

%%%%%%%%%%%%%%%%%%%%%%%%%%%%%%%%%%%%%%%%%%%%%%%%%%%%%%%%%%%%%%%%%%%%%%%
\subsubsection[Atypical unitary
and finite dimensional representations of the $\rm su(1,1|2)$ superalgebra]
{Atypical unitary and finite dimensional representations
\protect \\ of the $\bf su(1,1|2)$ superalgebra}\hspace*{\parindent}
\nopagebreak%%%%%%%%%%%%%%%%%%%%%%%%%%%%%%%%%%%%%%%%%%%%%%%%%%%%%%%%%%%%
Using equation (\ref{4.5}), one can now obtain the operators which
correspond to the Hamiltonian generators \usebox{\innhamnuber} of holomorphic
transformations of N=2 superunit disc. One gets
\begin{displaymath}
\begin{array}{l}\displaystyle
{\widehat J}{}_a=-\bar\xi_a\bar\partial- (\bar\partial\bar\xi_a)\left(
s+\frac12\bar\theta\frac{\partial}{\partial\bar\theta}
+\frac12\bar\chi\frac{\partial}{\partial\bar\chi}\right)\qquad{\widehat Z}=
s{\hat{\rm I}}
\end{array}
\end{displaymath}
\begin{equation}
\begin{array}{ll}\displaystyle
{\widehat P}_1=-\frac{1}{\sqrt{1-b^2}}\left(\frac{1-b}{2}\bar\chi
\frac\partial{\partial\bar\theta}
+\frac{1+b}{2}\bar\theta\frac\partial{\partial\bar\chi}\right) &\displaystyle
{\widehat P}_3=\frac12\bar\theta\frac\partial{\partial\bar\theta}
-\frac12\bar\chi\frac\partial{\partial\bar\chi} \\ \displaystyle
{\widehat P}_2=\frac{i}{\sqrt{1-b^2}}\left(\frac{1-b}{2}\bar\chi
\frac\partial{\partial\bar\theta}
-\frac{1+b}{2}\bar\theta\frac\partial{\partial\bar\chi}\right) &\displaystyle
{\widehat P}_4=\frac12\bar\theta\frac\partial{\partial\bar\theta}
+\frac12\bar\chi\frac\partial{\partial\bar\chi}
\end{array}\label{4.8}
\end{equation}
\begin{displaymath}
\begin{array}{l}\displaystyle
{\widehat E}{}^\alpha=\frac{\sqrt{1+b}}{2}\bar\theta\left[\bar{z}^\alpha\bar
\partial
+(\bar\partial\bar{z}^\alpha)\left(2s+\bar\chi\frac\partial{\partial\bar\chi}
\right)\right]
-\frac1{\sqrt{1+b}}\bar{z}^\alpha\frac\partial{\partial\bar\theta}
\\ \displaystyle
{\widehat F}{}^\alpha=
-i\frac{\sqrt{1+b}}{2}\bar\theta\left[\bar{z}^\alpha\bar\partial
+(\bar\partial\bar{z}^\alpha)\left(2s+\bar\chi\frac\partial{\partial\bar\chi}
\right)\right]
-i\frac1{\sqrt{1+b}}\bar{z}^\alpha\frac\partial{\partial\bar\theta}
\\ \displaystyle
{\widehat G}{}^\alpha=\frac{\sqrt{1-b}}{2}\bar\chi\left[\bar{z}^\alpha\bar
\partial
+(\bar\partial\bar{z}^\alpha)\left(2s+\bar\theta\frac\partial{\partial\bar
\theta}\right)\right]
-\frac1{\sqrt{1-b}}\bar{z}^\alpha\frac\partial{\partial\bar\chi}
\\ \displaystyle
{\widehat
H}{}^\alpha=-i\frac{\sqrt{1-b}}{2}\bar\chi\left[\bar{z}^\alpha\bar\partial
+(\bar\partial\bar{z}^\alpha)\left(2s+\bar\theta\frac\partial{\partial\bar
\theta}\right)\right]
-i\frac1{\sqrt{1-b}}\bar{z}^\alpha\frac\partial{\partial\bar\chi}\,,
\end{array}
\end{displaymath}
where $\bar\partial\equiv\partial/\partial\bar{z}$, all the derivatives
are left and $\bar\xi_a,\bar z^\alpha$ are defined by Eqs.~(\ref{3.16})
and~(\ref{3.7}) respectively. It is readily verified that the derived
operators generate an irreducible representation of~$\rm su(1|1|2)$
superalgebra, and it is the same case for any values $s$ and~$b$, not only
for $s>0$ and~$|b|<1$. The anticommutation relations for
operators~(\ref{4.8}) completely correspond to the Poisson
superbrackets~(\ref{3.22}), and it is sufficient to apply the
correspondence rules, that is, to replace $\{\,,\,\}\to1/i[\,,\,]_\mp$
(anticommutator for two odd operators and commutator in the rest cases) in
Rel.~(\ref{3.22}).  By reduction to the orthosymplectic subsuperalgebra we
reproduce just the typical UIR's of the $\rm osp(2|2)$ obtained
in~Refs.~\cite{Balant,Grad}.

The constructed representation is infinite dimensional for $s>0$,
$|b|<1$ and unitary in the sense that the operators (\ref{4.8}) are Hermitian
with respect to inner product \usebox{\scalprod}. It means, in particular,
that $\langle f|{\widehat J}{}_a|g\rangle_{{\cal L}^{1|2}}=\overline{\langle
g|{\widehat J}{}_a|f\rangle}_{{\cal L}^{1|2}}$, $\langle f|{\widehat
P}{}_I|g\rangle_{{\cal L}^{1|2 }}=\overline{\langle g|{\widehat
P}{}_I|f\rangle}_{{\cal L}^{1|2}}$ for any~${f,g\in{\cal O}_{s,b}}$. The
Hermitian selfconjugation conditions for the odd operators may reveal some
subtlety. Any odd classical observable among \usebox{\innhamnuber} is the
Majorana spinor and we have, for example, $\overline{E^0}=-E^1$,
$\overline{E^1}=-E^0$ with respect to the reality condition (\ref{Major}).
${\widehat E}{}^\alpha$~(and any odd
operator with the spinor index) is Hermitian in the sense that $\langle
f|{\widehat E}{}^0|g\rangle_{{\cal L}^{1|2}}= -\overline{\langle g|{\widehat
E}{}^1|f\rangle}_{{\cal L}^{1|2}}$.

We denote the UIR obtained by ${\bf D}_+^{s,b}$. With respect to the $\rm
su(1,1)$ subalgebra, it is decomposed into the direct sum $D^s_+\bigoplus
D^{s+1/2}_+\bigoplus D^{s+1/2}_+\bigoplus D^{s+1}_+$ of the unitary
representations of discrete series, and the components $f_0,f_{1,2},f_3$ of
the state (\ref{4.1}) transform by the representations of higher weights
$s,s+1/2$ and $s+1$ respectively.

The representations being obtained for $s\leq0$ or $|b|>1$ are non-unitary.
The case of $s+1=-j$, $j$ is non-negative integer or half integer, is special.
Then the operators (\ref{4.8}) generate a finite dimensional representation
${\bf D}^j$ of dimension $8j+8$. It is a superquartet of finite
dimensional representations of $\rm su(1,1)$, ${\bf D}^j=D^{j+1}\bigoplus
D^{j+1/2} \bigoplus D^{j+1/2}\bigoplus D^{j}$, and the state's components
$f_0,f_{1,2}, f_3$ transform by the $2j+3, 2j+2$ and $2j+1$ dimensional
representations respectively.

It  should be mentioned that the representations of the $\rm su(1,1|2)$ being
considered here correspond to an {\it irregular\/} coadjoint orbit ${\cal
L}^{1|2}$ of the supergroup $\rm SU(1,1|2)$ and, hence, they are {\it
atypical\/} representations. By reduction to the orthosymplectic
subsuperalgebra we get just the {\it typical\/} representations of the~$\rm
osp(2|2)$.

Keeping in mind the spinning superparticle we remember that the
representations~$D^s_+$ of the universal covering $\rm\overline
{SO^\uparrow(1,2)}$ are commonly used for conventional realizations of the
UIR's of D=3 Poincar\'e symmetry of fractional
spin~\cite{JackNair,Forte,CorPly}, whereas the finite dimensional irreps
$D^j$ serve the ones of integer or of half integer spin.  It will be natural
to extend these realizations to N=2 Poincar\'e supersymmetry by means of
representations ${\bf D}^{s,b}_+$ and ${\bf D}^j$ of the inner $\rm
su(1,1|2)$ superalgebra.

\subsubsection{The correspondence principle}
\hspace*{\parindent}\nopagebreak
%%%%%%%%%%%%%%%%%%%%%%%%%%%%%%%%%%%%%%%%%%%%%%%%%%%%%%%%%%%%%%%%%%%%%%%
To complete the quantization procedure on ${\cal L}^{1|2}$ let us return to
the relation between observables and linear operators. We have examined for
the supersymmetry generators of $\rm su(1,1|2)$ superalgebra that there is an
exact correspondence between supercommutators of the operators in ${\cal O}_{s,b}$
and the Poisson superbrackets of respective classical observables. In this
sense we do have ``the quantization'' of a classical mechanics on the N=2
superunit disc.  Consider now the correspondence between the algebras of
{\it arbitrary\/} linear operators in the Hilbert space and their symbols.

The problem we concern with is thoroughly studied for K\"ahler homogeneous
manifolds. Berezin proved the general ``correspondence
principle''~\cite{Berezin1,Berezin2}, which roughly consists in the
following. The multiplication of operators induces a binary $\ast$-operation
for corresponding symbols; $\ast$-multiplication is noncommutative.
Furthermore, the theory contains a ``Planck constant'' $h$ related to one of
the quantum numbers, and in the limit when $h\to0$ $\ast$-algebra transforms
to the ordinary commutative algebra of functions on the manifold. Finally,
the first order reset with respect to $h$ of the commutator of symbols
coincides with their Poisson bracket. The Lobachevsky plane has originally
served as a test example for the Berezin technique~\cite{Berezin2}. The
parameter~$s^{-1}$ plays the role of the Planck constant.

Similar principles hold true for N=2 superunit disc being
a natural N=2 superextension of the Lobachevsky plane.

Let ${\widehat A}{}_1,{\widehat A}{}_2$ be two linear operators in
${\cal O}_{s,b}$ and
$A_1({\mit\Gamma},\bar{\mit\Gamma}),A_2({\mit\Gamma},\bar{\mit\Gamma})$ being
the respective Berezin covariant symbols. It follows from Eq.~(\ref{4.5})
that the symbol being corresponded to the product ${\widehat A}{}_2\cdot
{\widehat A}{}_1$ (and denoted by $A_2\ast A_1$) reads
\begin{equation}
A_2\ast A_1({\mit\Gamma},\bar{\mit\Gamma})=\int_{{\cal L}_{1|2}}
A_2({\mit\Gamma}_1,\bar{\mit\Gamma})A_1({\mit\Gamma},\bar{\mit\Gamma}_1)
\frac{L_{s,b}({\mit\Gamma},\bar{\mit\Gamma}_1)L_{s,b}({\mit\Gamma}_1,
\bar{\mit\Gamma})}{L_{s,b}({\mit\Gamma}_1,\bar{\mit\Gamma}_1)L_{s,b}
({\mit\Gamma},\bar{\mit\Gamma})}d\mu({\mit\Gamma}_1,\bar{\mit\Gamma}_1)\,.
\label{4.9}
\end{equation}
Hence the multiplication of the operators induces the $\ast$-multiplications
of the symbols.

{\bf Theorem} (the correspondence principle): The following estimations
take place:
\begin{displaymath} \begin{array}{ll}\displaystyle 1)&\displaystyle
\lim\limits_{s\to\infty}A_2\ast A_1({\mit\Gamma},\bar{\mit\Gamma})=
A_2({\mit\Gamma},\bar{\mit\Gamma})\cdot A_1({\mit\Gamma},\bar{\mit\Gamma})\\
\displaystyle 2)&\displaystyle
\lim\limits_{s\to\infty}s\left(A_2\ast A_1({\mit\Gamma},\bar{\mit\Gamma})
-A_1\ast A_2({\mit\Gamma},\bar{\mit\Gamma})\right)=is
\{A_2\,,\,A_1\}\,,
\end{array}
\end{displaymath}
where $\{\,,\,\}$ is the Poisson superbracket on ${\cal L}^{1|2}$.

To examine the correspondence principle we need for the explicit form of the
fundamental Poisson superbrackets $\Omega^{A\bar B}=\{\Gamma^A,\Gamma^{\bar
B}\}$. They are derived accounting for the condition $\Omega^{A\bar
B}\Omega_{\bar{B}C}= \delta^A{}_C$, where $\Omega_{\bar{A}B}$ is the
supermatrix of the symplectic two-superform~(\ref{3.10}), $\Omega_{{\cal
L}^{1|2}}\equiv{\rm d} {\mit\Gamma}^{\bar A}\Omega_{\bar AB}{\rm
d}{\mit\Gamma}^B$. A slightly cumbersome calculation leads to
\begin{eqnarray}\displaystyle &&
\{z,\bar z\}=-\frac{i}{2s}(1-z\bar z)^2
\left(1+\frac{1+b}{2}\frac{\theta\bar\theta}{1-z\bar z}
+\frac{1-b}{2}\frac{\chi\bar\chi}{1-z\bar z}\right)\nonumber\\
&&\displaystyle
\{z,\bar\theta^I\}=\frac{i}{2s}z\bar\theta^I(1-z\bar z)
\left(1+\frac{1+b}{2}\frac{\theta\bar\theta}{1-z\bar z}
+\frac{1-b}{2}\frac{\chi\bar\chi}{1-z\bar z}\right)\quad
\nonumber\\[1mm]&&\displaystyle
\{\theta^I,\bar z\}=\frac{i}{2s}\bar z\theta^I(1-z\bar z)
\left(1+\frac{1+b}{2}\frac{\theta\bar\theta}{1-z\bar z}
+\frac{1-b}{2}\frac{\chi\bar\chi}{1-z\bar z}\right)\label{4.11}\\
&&\displaystyle
\{\theta,\bar\theta\}=-\frac{i}{s}(1-z\bar z)
\left(\frac1{1+b}+\frac{1-b}{2(1+b)}\frac{\chi\bar\chi}{1-z\bar z}
+\frac{z\bar z}{2}\frac{\theta\bar\theta}{1-z\bar z}
-\frac{1-b}4\frac{\theta\bar\theta\chi\bar\chi}{1-z\bar z}\right)\nonumber\\
&&\displaystyle\{\chi,\bar\chi\}=-\frac{i}{s}(1-z\bar z)
\left(\frac1{1-b}+\frac{1+b}{2(1-b)}\frac{\theta\bar\theta}{1-z\bar z}
+\frac{z\bar z}{2}\frac{\chi\bar\chi}{1-z\bar z}
-\frac{1+b}4\frac{\theta\bar\theta\chi\bar\chi}{1-z\bar z}\right)\nonumber\\
&&\displaystyle\{\theta,\bar\chi\}=\frac{i}{2s}(1-z\bar z)\theta\bar\chi
\qquad\{\chi,\bar\theta\}=\frac{i}{2s}(1-z\bar z)\chi\bar\theta\,,\nonumber
\end{eqnarray}
where $\theta^I\equiv(\theta,\chi)$. It is seen, in particular, that the
r.h.s. of the fundamental Poisson superbrackets contains the order $s^{-1}$.

{\bf Proof}: It is based on the asymptotic estimation
\begin{equation}
A_2\ast A_1({\mit\Gamma},\bar{\mit\Gamma})=
A_2({\mit\Gamma},\bar{\mit\Gamma})\cdot A_1({\mit\Gamma},\bar{\mit\Gamma})
+iA_2({\mit\Gamma},\bar{\mit\Gamma})\frac{\stackrel{\leftarrow}
{\partial}\phantom{0}}{\partial{\mit\Gamma}^A}\Omega^{A\bar B}
\frac{\stackrel{\rightarrow}{\partial}\phantom{0}}
{\partial{\mit\Gamma}^{\bar B}}A_1({\mit\Gamma},\bar{\mit\Gamma})
+{\cal O}(s^{-2})\,,\label{4.12}
\end{equation}
from which both propositions of the theorem are easily obtained. The
validity of the latter relation is sufficient to prove when $z=0$. If it
is the case, Eqs.~(\ref{4.12}) hold true at any $z$ in consequence of the $\rm
SU(1,1)$ invariance of the symplectic structure. Taking this fact into
account the verification of Eq.~(\ref{4.12}) is made by means of an ordinary
expansion of the symbols in (finite) series in the odd variables and the
comparison of l.h.s.\ and r.h.s.\ of Eqs.~(\ref{4.12}) for the
respective components.  It is a trivial but cumbersome exercise, which may be
successfully performed using the known estimation~\cite{Berezin2}
\begin{equation}
{\widehat T}{}_l[\varphi]\equiv\frac{2l-1}{2\pi i}\int\limits_{|z|<1}
\varphi(z,\bar z)(1-z\bar z)^{2l-2}dzd\bar z=\varphi(0,0)+\frac{1}{2l}\left.
\bigtriangleup\varphi(z,\bar z)\right|_{\scriptstyle z=\bar z=0}
+{\cal O}(l^{-2})\label{4.13}
\end{equation}
and its consequence
\begin{displaymath}
l({\widehat T}{}_l[\varphi]-
{\widehat T}{}_{l+1/2}[\varphi])=\frac1{4l}\left.\bigtriangleup
\varphi(z,\bar z)\right|_{\scriptstyle z=\bar z=0}+{\cal O}(l^{-2})\,.
\end{displaymath}
Here $l>1/2$, $\varphi(z,\bar z)$ is an arbitrary function to be continuously
differentiable into the unit disc in a complex plane, and $\bigtriangleup=
(1-z\bar z)^2\partial\bar\partial$ is an invariant Laplace-Beltrami operator
in ${\cal L}$. It is exactly the estimation (\ref{4.13}),
which has been originally applied by Berezin for the proof of the
correspondence principle in the Lobachevsky plane~\cite{Berezin2}. In this
sense, we reduce the correspondence principle in ${\cal L}^{1|2}$  to the
one in ${\cal L}$ by means of the expansion in the odd variables.
\rule{2mm}{2mm}

%%%%%%%%%%%%%%%%%%%%%%%%%%%%%%%%%%%%%%%%%%%%%%%%%%%%%%%%%%%%%%%%%%%%%%%%
\subsection[Operator realization of the Poincar\'e superalgebra.
Renormalization of the supercharges] {\normalsize\bf\hspace{-2mm}Operator
realization of the Poincar\'e superalgebra. \protect\\
\hspace*{-2mm}Renormalization of the supercharges} \label{sbs4.2}
\hspace*{\parindent}\nopagebreak
%%%%%%%%%%%%%%%%%%%%%%%%%%%%%%%%%%%%%%%%%%%%%%%%%%%%%%%%%%%%%%%%%%%%%%%%%
Now we are in a position to proceed directly to the quantization of D=3
spinning superparticle. Consider the space ${\cal H}$ of functions of the form
\begin{eqnarray}\displaystyle
F(p,\bar{\mit\Gamma})\equiv
F(p,\bar{z},\bar{\theta},\bar{\chi})&\displaystyle =& \displaystyle
F_0(p,\bar{z})+\sqrt{s(1+b)}\,\bar\theta F_1(p,\bar{z})
+\sqrt{s(1-b)}\,\bar\chi F_2(p,\bar{z})\nonumber\\&&\displaystyle
+\sqrt{s(s+1/2)(1-b^2)}\,\bar\theta\bar\chi F_3(p,\bar{z})\,,\label{4.14}
\end{eqnarray}
where $p\equiv p^a\in{\rm R}^{1,2}$, and $F_p(\bar{\mit\Gamma})\equiv
F(p,\bar{\mit\Gamma})\in{\cal O}_{s,b}$ at each fixed $p$. We would like
suppose that the Hamiltonians (\ref{3.23}) (which are the same as in Rel.\
(\ref{3.12})) present ``the classical symbols'' of respective operators of
the N=2 Poincar\'e superalgebra acting in~${\cal H}$. We take
the following ansatz for these operators
\begin{equation}
\begin{array}{l}\displaystyle
{\widehat{\cal J}}{}_a=-i\epsilon_{abc}p^b\frac\partial{\partial p_c}
+{\widehat J}{}_a\qquad\qquad {\widehat{\cal P}}_a=p_a\qquad\qquad
{\widehat{\cal Z}}=mb\\ \displaystyle
{\widehat{\cal Q}}{}^1_\alpha=(ip_{\alpha\beta}{\widehat
W}^\beta+m{\widehat{\tilde W}}{}_\alpha)[1+{\rm q}(b{\widehat
P}{}_3-\sqrt{1-b^2}\,{\widehat P}{}_2-{\widehat P}{}_4)] \\ \displaystyle
{\widehat{\cal Q}}{}^2_\alpha
=(ip_{\alpha\beta}{\widehat
V}^\beta+m{\widehat{\tilde V}}{}_\alpha) [1+{\rm q}(b{\widehat
P}{}_3+\sqrt{1-b^2}\,{\widehat P}{}_2-{\widehat P}{}_4)]\,.
\end{array} \label{4.15}
\end{equation}
Here the operators ${\widehat
W}^\alpha,{\widehat{\tilde W}}{}^\alpha,{\widehat V}^\alpha, {\widehat{\tilde
V}}{}^\alpha$ are expressed as linear combinations of ${\widehat
E}{}^\alpha,{\widehat F}{}^\alpha$, ${\widehat G}{}^\alpha$, ${\widehat
H}{}^\alpha$ according to relations (\ref{3.24}), whereas the latter, together
with the operators~${\widehat J}{}_a$ and~${\widehat P}{}_I$, are defined
by the expressions (\ref{4.8}).

Recall that the classical observables (\ref{3.23}) or (\ref{3.12}) generate
the Poincar\'e superalgebra on shell only, that is modulo to the constraints
(\ref{3.14}). The operator counterparts of the constraints are now imposed
to annihilate the physical states according to Dirac quantization
prescriptions. The linear operator in ${\cal O}_{s,b}$, which corresponds to
the Berezin covariant symbol $-sn_a$, ($s>0$) reads
\begin{displaymath}
\widehat{-sn_a}={\widehat J}{}_a\left(1-\frac{2}{2s+1}{\widehat
P}{}_4+\frac{2}{(2s+1)(2s+2)}
\bar\theta\frac{\stackrel\rightarrow\partial}{\partial\bar\theta}
\bar\chi\frac{\stackrel\rightarrow\partial}{\partial\bar\chi}\right)=
{\widehat J}{}_a\left(1+\frac1s\widehat P_4\right)^{-1}\,.
\end{displaymath}
Thus, the wave equations for the superparticle are easily brought to
the form\footnote{It is worth noting that the second constraint equation
(\ref{3.14}) should be written in an equivalent form $(p,J)-mP_4-ms=0$.}
\begin{equation}
\begin{array}{l}\displaystyle
(p^2+m^2)F^{\rm phys}(p,\bar z,\bar\theta,\bar\chi)=0\\ \displaystyle
[(p,{\widehat J})-m{\widehat P}{}_4-ms]F^{\rm phys}(p,\bar
z,\bar\theta,\bar\chi)=0\,.
\end{array} \label{4.16}
\end{equation}
Solutions of the wave equations generate a subspace ${\cal H}_{m,s,b}$ in
${\cal H}$. Furthermore, if~\mbox{$F\in{\cal H}_{m,s,b}$} and
${\widehat{\cal S}}$ is any one of the operators (\ref{4.15}), then
${\widehat{\cal S}}F$ is a physical state again, regardless of the
particular value of the parameter $\rm q$. In this sense, the wave
equations are superpoincar\'e invariant.

It is crucial now to examine explicitly whether the operators
(\ref{4.15}) actually generate the N=2 Poincar\'e superalgebra. One gets
in ${\cal H}$ (compare with Eq.\ (3.13))
\begin{equation}
\begin{array}{lll}\displaystyle
[{\widehat{\cal J}}{}_a,{\widehat{\cal
J}}{}_b]_-=i\epsilon_{abc}{\widehat{\cal J}}{}^c & \displaystyle
[{\widehat{\cal J}}{}_a,{\widehat{\cal P}}{}_b]_-=i\epsilon_{abc}
{\widehat{\cal P}}{}^c & \displaystyle
[{\widehat{\cal P}}{}_a,{\widehat{\cal P}}{}_b]_-=0\\ \displaystyle
[{\widehat{\cal J}}{}_a,{\widehat{\cal Q}}{}^I_\alpha]_-
=\frac{1}{2}(\gamma_a)_\alpha{}^\beta
{\widehat{\cal Q}}{}^I_\beta &\displaystyle [{\widehat{\cal P}}{}_a,
{\widehat{\cal Q}}{}^I_\alpha]_-=0&\displaystyle
[{\widehat{\cal Q}}{}^I_\alpha,{\widehat
P}_3]_-=-\frac{i}{2}\epsilon^{IJ}{\widehat{\cal Q}}{}^J_\alpha\,.
\end{array}\label{4.17}
\end{equation}
These relations hold true with an arbitrary number taken for {\rm q}.
However, the anticommutator of the supercharges is strongly dependent on
the particular value of~${\rm q}$:
\begin{equation}
[{\widehat{\cal Q}}{}^I_\alpha,{\widehat{\cal Q}}{}^J_\beta]_+=
2\delta^{IJ}p_{\alpha\beta}-2imb\epsilon^{IJ}\epsilon_{\alpha\beta}
+c^{1\,}{}^{IJ}_{\alpha\beta}(p^2+m^2)+c^{2\,}{}^{IJ}_{\alpha\beta}
((p,{\widehat J})- m{\widehat P}{}_4-ms)+{\cal O}(s^{-2})\,,\label{4.18}
\end{equation}
where $c^{\,\cdot\,}{}^{IJ}_{\alpha\beta}$ are some functions, and ${\cal
O}(s^{-2})$ are the corrections of higher orders in $s^{-1}$, which depend on
$\rm q$ and the other parameters like $m$ and $b$. These corrections do
not vanish if ${\rm q}={\rm q}^{cl}=1/4s$.

Eq.~(\ref{4.18}) is presented in more detailed notation in the
Appendix~(Eqs.~(\ref{A5})) from the view of the closing of the operator
superalgebra.

Notice that the quantum value of ${\rm q}$ is not
uniquely determined. The value \mbox{${\rm q}^{cl}=1/4s$} is derived from
the expressions (\ref{3.23}), when the relationship between the
superpoincar\'e and $\rm su(1,1|2)$ generators \usebox{\innhamnuber} is
taken into account.  However, one can start immediately from the symbols
(\ref{3.12}) and restore the operators applying the correspondence
rule~(\ref{4.5}). What is remarkable is that one obtains the same
operators~(\ref{4.15}), but the parameter ${\rm q}$ changes, and ${\rm
q}^{cl}$ appears to be ${\rm q}^{cl}_1=1/(4s+2)$. But the Poincar\'e
superalgebra is disclosed by ${\rm q}=1/(4s+2)$;  the same is true
for~\mbox{${\rm q}=1/4s$} too.

Both the appearance of corrections ${\cal O}(s^{-2})$ in r.h.s. of
Eq.~(\ref{4.18}) and the ambiguity in the definition of ${\rm q}$ have
the same origin. That is, a {\it nonlinearity\/} of the Poincar\'e supercharge
operators (\ref{4.15}) in the generators (\ref{4.8}) of the inner
$\rm su(1,1|2)$ superalgebra. In consequence of the nonlinearity, different
operator factor orderings may lead to the different forms for ${\widehat{\cal
Q}}{}^I_\alpha$, and the corrections appear in response to the correspondence
principle in ${\cal L}^{1|2}$.

We show that the disclosure of the Poincar\'e superalgebra at the quantum
level has transparent mathematical ground in view of the Berezin
correspondence principle. However, this disclosure is quite unsatisfactory
from the physical viewpoint for the quantization of the elementary system. The
latter is completely characterized by its inherent symmetries (in the present
case it is the D=3 Poincar\'e SUSY). {\it It is the representation
of these symmetries in Hilbert space, which allows to identify the obtained
quantum theory with the quantized elementary system.\/} According to
these reasons, to quantize D=3 superparticle we now have to provide an
{\it exact\/} realization of the representation of the
Poincar\'e superalgebra in
the physical Hilbert space, without any corrections in the parameters of the
model. To find the true quantum realization for the representation, we can try,
starting from Eqs.~(\ref{4.15})\,--\,(\ref{4.18}), to introduce some
renormalized terms in the observables (\ref{4.15}), which should be
sufficient for the closure of the anticommutators (\ref{4.18}).

Certainly, we don't have an a priori reason, which may ensure the
consistency of the renormalization procedure; a structure of the possible
higher order corrections to~(\ref{4.15}) is unclear in general. Surprisingly,
the exact corrections may be obtained from the simplest ansatz (\ref{4.15}) for
the quantum observables. In other words, a true ordering exists for the
$\rm su(1,1|2)$ superalgebra operators, entering in ${\widehat{\cal
Q}}{}^I_\alpha$ in Eq.~(\ref{4.15}), that {\it allows to restore a
representation of the Poincar\'e superalgebra by the renormalization of
the only parameter ${\rm q}$\/}.

It is examined by a direct calculation that the corrections ${\cal O}(s^{-2})$
in r.h.s.\ of Eqs.~(\ref{4.18}) vanish and the operators (\ref{4.15})
 generate the closed Poincar\'e superalgebra if, and only if
\begin{equation}
{\rm q}={\rm q}^{quant}_\mp=1\mp\sqrt{1-\frac{1}{2s+1}}\ .
\label{4.19}
\end{equation}
Some details of calculations of the anticommutators (\ref{4.18}) are given
in the Appendix. The renormalized value ${\rm q}^{quant}_-={\rm q}^{cl}+{\cal
O}(s^{-2})$ can be treated as a perturbative correction to the classical
symbols of the supercharges. The other possible value
${\rm q}^{quant}_+$ emerges from the hidden N=4 supersymmetry and could
be understood  from the following reasons.

Let ${\rm q=q_-}\,$. The operators of supercharges corresponding to the
classical observables~(\ref{N4}) (see also Eqs.~(\ref{relN4}),~(\ref{3.12a}))
and providing the hidden N=4 supersymmetry in ${\cal H}_{m,s,b}$ are
presented by
\begin{equation}
{\widehat{\tilde{\cal Q}}}{}^I_\alpha
=-\frac{i}{m}p_\alpha{}^\beta\widehat{\cal Q}^I_\beta=
\left.i\widehat{K}\cdot\widehat{\cal Q}^I_\alpha\right|_{{\rm q}\to2-{\rm q}}
\,,
\label{N4op}
\end{equation}
where the {\it parity operator} $\widehat{K}$ is introduced. It acts on the
components of the wave function (\ref{4.14}) by the rule
\begin{equation}
\widehat{K}:F=(F_0,F_1,F_2,F_3)\to\widehat{K}F=(F_0,-F_1,-F_2,F_3)\qquad
\widehat{K}^2=\hat{\rm I}\,, \label{parityop}
\end{equation}
and $\left.\widehat{\cal Q}^I_\alpha\right|_{{\rm q}\to2-{\rm q}}$ denote
the supercharges (\ref{4.15}) being considered when the constant ${\rm q}$
is substituted for the expression $(2-{\rm q})$.

The same critical values (\ref{4.19}) evidently provide the closure of
the N=4 Poincar\'e superalgebra. Moreover, the parity operator (\ref{parityop})
possesses remarkable features: it commutes with the even generators of the
N=4 Poincar\'e superalgebra and anticommutes with the supercharges:
\begin{equation}
\begin{array}{c}\displaystyle
[{\widehat{\cal
J}}{}_a,\widehat{K}]_-=[\widehat{\cal P}{}_a,\widehat{K}]_-=0 \qquad
[\widehat{\cal P}{}_I,\widehat{K}]_-=0\quad (I=1,2,3,4)\\
\displaystyle
[{\widehat{\cal Q}}{}^I_\alpha,{\widehat K}]_+=
[\widehat{\tilde{\cal Q}}{}^I_\alpha,{\widehat K}]_+=0\quad I=1,2\,.
\end{array}\label{paritycomm}
\end{equation}
Therefore, the operators $\widehat{{\cal Q}}{}^{\prime}{}^I_\alpha=
-i\widehat{K}\widehat{{\cal Q}}^I_\alpha=
\widehat{\tilde{\cal Q}}{}^I_\alpha|_{{\rm q}\to2-{\rm q}}$ and
$\widehat{\tilde{\cal Q}}{}^{\prime}{}^I_\alpha=-i\widehat{K}
\widehat{\tilde{\cal Q}}{}^I_\alpha
=\widehat{\cal Q}{}^I_\alpha|_{{\rm q}\to2-{\rm q}}$ satisfy
the (anti)commutation relations being identical with  Eqs.~(\ref{4.17}),
(\ref{4.18}) for the supercharges $\widehat{\cal Q}^I_\alpha$ and
${\widehat{\tilde{\cal Q}}}{}^I_\alpha$ themselves. This observation
clarifies to some extent the origin of the nonperturbative value  ${\rm q}^{quant}_+$
for the parameter ${\rm q}$. Notice that two representations of the
Poincar\'e superalgebra corresponding to either possible value $\rm q$
are equivalent to each other. It is seen straightforwardly from relations
\begin{displaymath}
\begin{array}{c} \widehat{\cal
Q}{}^{\prime}{}^I_\alpha=\widehat U \widehat{\tilde{\cal Q}}{}^I_\alpha
\widehat U\qquad \widehat{\tilde{\cal
Q}}{}^{\prime}{}^I_\alpha=\widehat U \widehat{\cal
Q}{}^I_\alpha\widehat U \qquad
[{\widehat{\cal J}}_a, \widehat U]_-=[{\widehat{\cal P}}_a, \widehat U ]_-
=[\widehat P_I, \widehat U ]_-=0\,,
\end{array}
\end{displaymath}
where the operator $\widehat U$ reads
\begin{displaymath}
\widehat U=1-2\bar\theta\frac{\stackrel\rightarrow\partial}{\partial\bar
\theta}\bar\chi\frac{\stackrel\rightarrow\partial}{\partial\bar\chi}\qquad
\qquad\widehat U^2=\hat{\rm I}\,.
\end{displaymath}

We don't observe, however, either any  classical counterpart for
the supercharges $\widehat{\cal Q}{}^{\prime}{}^I_\alpha\,,
\widehat{\tilde{\cal Q}}{}^{\prime}{}^I_\alpha$ or any
algebraic construction (for instance, superalgebra) involving both sets
of the N=4 supercharges $\widehat{\cal Q}^I_\alpha\,, \widehat{\tilde{\cal
Q}}{}^I_\alpha$ and $\widehat{\cal Q}{}^{\prime}{}^I_\alpha\,,
\widehat{\tilde{\cal Q}}{}^{\prime}{}^I_\alpha$ on equal footing.

To summarize briefly, the ``double'' SUSY of the classical mechanics of D=3
spinning superparticle can be lifted to the operator representation in the
quantum theory. The key step of construction is the renormalization
(\ref{4.19}) for the Poincar\'e supercharges (\ref{4.17}). Eq.~(\ref{4.19})
displays two exceptional values of the parameter ${\rm q}$ providing the
closure for the anticommutator of supercharges (\ref{4.18}) and recovering
the consistent representation of the Poincar\'e superalgebra. We will suppose
below that the parameter ${\rm q}$ is equal to either of two ${\rm
q}^{quant}_\mp$.  We will show below, that the
representation space ${\cal H}_{m,s,b}$ is endowed by a natural Hilbert space
structure, thus the operators (\ref{4.14}) and (\ref{N4op}) of the
Poincar\'e representation become Hermitian. To be specific, we will
consider explicitly the operators (\ref{4.14}) only, which give a
representation of N=2 Poincar\'e superalgebra. The appearance of the hidden N=4
supersymmetry, being presented by additional supercharges (\ref{N4op}),
becomes thereby obvious.

%%%%%%%%%%%%%%%%%%%%%%%%%%%%%%%%%%%%%%%%%%%%%%%%%%%%%%%%%%%%%%%%%%%%%%%%
\subsection[Hilbert space for N=2 superanyon]
{Hilbert space for N=2 superanyon} \label{sbs4.3}
\hspace*{\parindent}\nopagebreak
%%%%%%%%%%%%%%%%%%%%%%%%%%%%%%%%%%%%%%%%%%%%%%%%%%%%%%%%%%%%%%%%%%%%%%%%%
For an arbitrary $s>0$ and $|b|<1$ the physical space ${\cal H}_{m,s,b}$ is
naturally endowed by an inner product
\begin{equation}
(F|G)={\cal N}\int\frac{d{\bf p}}{p^0}\langle F|G\rangle_{{\cal L}^{1|2}}
\qquad p^0=\sqrt{{\bf p}^2+m^2}>0\,,\label{4.20}
\end{equation}
where $\langle F|G\rangle_{{\cal L}^{1|2}}$ denotes the inner product
\usebox{\scalprod} in ${\cal O}_{s,b}\, $, $p^a=(p^0,{\bf p})$ and ${\cal N}$
is an arbitrary normalization constant. The operators~(\ref{4.15}) of the N=2
Poincar\'e superalgebra are Hermitian with respect to the introduced inner
product.  This fact follows immediately from that the Hermitian property
for the  operators of the inner
$\rm su(1,1|2)$ superalgebra with respect to
$\langle\cdot|\cdot\rangle_{{\cal L}^{1|2}}$ and
\begin{displaymath}
[{\widehat S}^{I\,\alpha}_\pm\,,\,b{\widehat P}{}_3\pm\sqrt{1-b^2}\,{\widehat
P}{}_2-{\widehat P}{}_4]_- =0 \qquad{\widehat S}^{I\,\alpha}_+\equiv({\widehat
V}{}^\alpha,{\widehat{\tilde V}}{}^\alpha) \qquad {\widehat
S}{}^{I\,\alpha}_-\equiv({\widehat W}{}^\alpha,{\widehat{\tilde
W}}{}^\alpha)\,
\end{displaymath}
(compare with the latter notion in Subsec.~III.7).

Thus, we have constructed a unitary representation of N=2, D=3 superalgebra
in the Hilbert space ${\cal H}_{m,s,b}$. In terms of the expansion
(\ref{4.14}), the wave equations (\ref{4.16}) for components reduce to
\begin{equation}
(p^2+m^2)F_\imath^{\rm phys}(p,\bar z)=0\qquad
[(p,{\widehat J}{}^l)-ml]F_\imath^{\rm phys}(p,\bar z)=0\,,
\label{4.21}
\end{equation}
where $i=0,1,2,3$;\hspace{3mm} $l=s$ for $i=0$,\hspace{3mm} $l=s+1/2$ for
$i=1,2$,\hspace{3mm} $l=s+1$ for $i=3$ and~${\widehat
J}{}^{\,l}_a= -\bar\xi_a\bar\partial-l(\bar\partial\bar\xi_a)$;
$\bar\xi_a$ is defined by Eq.~(\ref{3.16}). It is well
known~\cite{JackNair,CorPly,GKL2} that the solutions of
equations~(\ref{4.21}) generate the physical Hilbert space of D=3 particle of
mass $m$, arbitrary fixed fractional spin $l>0$ and positive energy $p^0>0$.
In this realization the fields $F_i(p,\bar z)$ carry an infinite dimensional
UIR's $D^l_+$ of the group~$\rm\overline{SO^\uparrow(1,2)}$.

Now, we may conclude that the supersymmetric theory describes the irreducible
superquartet of anyons of mass $m$, superspin $s$ and central charge $|b|<1$.
The corresponding representation is realized on the fields carrying the
atypical UIR ${\bf D}^{s,b}_+$ of the superalgebra $\rm su(1,1|2)$ (or the
typical one of $\rm osp(2|2)$).

It is worth  recalling that the Poincar\'e superalgebra admits unitary
representations, when $p^0>0$ and the central charge satisfies the
Bogomol'nyi-Prassad-Sommerfield bound $m\geq|{\cal Z}|$ (that is $|b|\leq1$
in our case) \cite{sohn}. In our realization, the wave
equations~(\ref{4.16}),~(\ref{4.21}) admit, in fact, the positive energy
solutions only, and the Hermitian conditions for operators (\ref{4.8}) and
(\ref{4.15}) are broken when $|b|>1$. In addition we observe a shortening of
the massive N=2 supermultiplet to a hypermultiplet in the BPS limit $|b|=1$;
the latter case will be briefly discussed in the last subsection.

%%%%%%%%%%%%%%%%%%%%%%%%%%%%%%%%%%%%%%%%%%%%%%%%%%%%%%%%%%%%%%%%%%%%%%%%%
\subsection[Hilbert space for N=2 superparticle of (half)integer
superspin]{Hilbert space for N=2 superparticle\protect\\ of
(half)integer superspin} \label{sbs4.4}\hspace*{\parindent}\nopagebreak
%%%%%%%%%%%%%%%%%%%%%%%%%%%%%%%%%%%%%%%%%%%%%%%%%%%%%%%%%%%%%%%%%%%%%%%%%
In contrast to the anyon case, the ordinary states carrying (half)integer spin
have conventional realization in terms of the {\it finite} component
spin-tensor fields in the Minkowski space. We have mentioned above that the
finite dimensional representations of the superalgebra $\rm su(1,1|2)$ may
emerge at $s+1=-j\,,2j=0,1,2, \dots$. These representations are non-unitary.
In particular, the inner product \usebox{\scalprod} (and,
thus,~(\ref{4.20})) has the inherent singularity at $|z|=1$, and the
consideration of the previous subsection becomes inadequate in this case.
We construct here a correct realization of the Hilbert space in the case
of (half)integer (super)spin, which enables to reproduce the conventional
description in terms of the spin-tensor fields.

Let us start from the spinning particle without supersymmetry living
classically in the phase space ${\cal M}^8=T^\ast({\rm R}^{1,2})\times
{\cal L}$, that is, the model of Sec.~II.
The Hilbert space of the particle of (half)integer spin $j>0$
can be realized in the space ${\cal H}_j$ of the fields on ${\cal M}^8$
of the following form:
\begin{equation} F(p,\bar
z)=F_{\alpha_1\alpha_2\dots\alpha_{2j}}(p){\bar z}^{\alpha_1} {\bar
z}^{\alpha_2}\dots{\bar z}^{\alpha_{2j}}\qquad \alpha_k=0,1\,,\label{4.22}
\end{equation}
where the coefficients $F_{\alpha_1\alpha_2\dots\alpha_{2j}}(p)$ are totally
symmetric in their indices. $F(p,\bar z)$ appears to be polynomial
of the degree $2j$ in the variable $\bar z$.

The following consideration is based on remarkable transformation properties
of the twistorlike objects $z^\alpha$ and ${\bar z}^\alpha$ defined by
Eqs.~(\ref{3.7}). The Lorentz group $\rm SO^\uparrow(1,2)\cong SU(1,1)/{\bf
Z}_2$ acts on $\cal L$ by fractional linear transformations
\begin{equation}
N:\;\;z\rightarrow
z^{\prime}=\frac{az-b}{\bar a-\bar{b} z} \qquad \qquad \|N_\alpha{}^\beta\|=
\left(\begin{array}{cc}a&b\\ \bar b&\bar a\end{array}\right)\in SU(1,1)\;,
\label{4.23}
\end{equation}
which may be rewritten identically as
\begin{equation}
N:\ z^{\alpha}\to z^{\alpha\,\prime} = \left(
\frac{\partial z^\prime}{\partial z}\right )^{1/2} N^{-1}{}_\beta {}^\alpha
z^\beta\qquad \bar{z}^\alpha\rightarrow \bar{z}^{\alpha\,\prime}= \left(
\frac{\partial \bar{z}^\prime}{\partial \bar{z}}\right )^{1/2} N^{-1}{}_\beta
{}^\alpha \bar{z}^\beta\;,
\label{4.24}
\end{equation}
or, in the infinitesimal form,
\begin{displaymath}
\delta z=\frac{i}{2}\omega_{\alpha\beta} z^{\alpha} z^{\beta}\qquad
\delta \bar{z}=-\frac{i}{2}\omega_{\alpha\beta} \bar{z}^{\alpha}
\bar{z}^{\beta}\,,
\end{displaymath}
where $\omega_{\alpha\beta}\equiv(\omega^a\gamma_a )_{\alpha\beta}$
are the parameters of infinitesimal Lorentz transformations. As is seen,
each of $z^\alpha$, ${\bar z}^\alpha$ transforms simultaneously as a D=3
Lorentz spinor and as field density in $\cal L$. They are, in fact, the
only independent twistorlike fields associated with the homogeneous space
structure on the Lobachevsky plane. We have in particular
\begin{equation}
n^{\alpha\beta}=n^a\gamma_{a}^{\alpha\beta}=-\frac{z^\alpha{\bar
z}^\beta+z^\beta{\bar z}^\alpha}{ (\epsilon_{\gamma\delta}z^\gamma{\bar
z}^\delta)}\qquad \Omega_{\cal L}=-2is\frac{{\rm d}z\wedge{\rm d}\bar z}{
(\epsilon_{\gamma\delta}z^\gamma{\bar z}^\delta)^2}\,.\label{4.25}
\end{equation}
for the unit timelike Lorentz vector $n^a$ (\ref{2.7}) and the K\"ahler
two-form (\ref{2.6}) in the Lobachevsky plane.

Let us suppose the coefficients $F_{\alpha_1\alpha_2\dots\alpha_{2j}}(p)$
in Eq.\ (\ref{4.22}) to be Lorentz spin-tensor field of the type
$j$. Then $F(p,\bar z)$ possesses the following transformation
law with respect to the action of Lorentz group
\begin{displaymath}
N:F(p,\bar z)\to F^\prime(p,{\bar z}^\prime)=\left(\frac{\partial{\bar
z}^\prime}{\partial{\bar z}}\right)^jF(p,{\bar z}) \,.
\end{displaymath}
We have a standard realization of the finite dimensional representation $D^j$
\cite{GelfGrVil,Perel,JackNair}. Extending this
construction to the representation of the Poincar\'e group in ${\cal H}_j$ one
takes the following transformation law of the fields
\begin{equation}
F(p,\bar z)\to
F^\prime(p^\prime,{\bar z}^\prime)=\left(\frac{\partial{\bar
z}^\prime}{\partial{\bar z}}\right)^jF(p,{\bar z})\,.\label{4.26}
\end{equation}
The Poincar\'e generators read
\begin{equation}
{\widehat{\cal P}}{}_a=p_a \qquad{\widehat{\cal J}}{}_a=-i\epsilon_{abc}p^b
\frac{\partial}{\partial p_c}+{\widehat J}{}^j_a\,,
\label{4.27}
\end{equation}
where
\begin{equation}
{\widehat J}{}^j_a=-\bar\xi_a{\bar\partial}+j({\bar\partial}{\bar\xi}_a)
\end{equation}
and we recall again that $\bar\xi_a$ is defined by Eq.~(\ref{3.16}).
Now, to separate  a subspace of irreducible representation
of the Poincar\'e group from ${\cal H}_j$ it is sufficient to impose the operator counterparts
of the constraints (\ref{2.11}) to annihilate the physical states:
\begin{equation}
(p^2+m^2)F(p,\bar z)=0\qquad [(p,{\widehat J}{}^j)+mj]F(p,\bar
z)=0\label{4.28}
\end{equation}
The space ${\cal H}_{m,j}$ of solutions of the wave equations (\ref{4.28})
generates the irreducible one- or two-valued massive representation of
$\rm ISO^\uparrow(1,2)$. Moreover, using the
identity~${\bar z}^\alpha\bar\partial{\bar z}^\beta-{\bar
z}^\beta\bar\partial{\bar z}^\alpha= \epsilon^{\alpha \beta}$ one can verify
that the irreducibility conditions are equivalent to the following set
of equations for Lorentz spin-tensors:
\begin{equation}
p_{\alpha_1}{}^{\beta}F_{\beta\alpha_2\dots\alpha_{2j}}(p)=
mF_{\alpha_1\alpha_2\dots\alpha_{2j}}(p)\label{4.29}\,.
\end{equation}
On-shell, the only independent component survives among $2j+1$ ones
(for instance, in the rest system, where $p^a=(m,0,0)$, the only nonvanishing
component is $F_{11\dots1}(p)$).

It is now easy to write down the well-defined Poincar\'e
invariant inner product
in the space ${\cal H}_{m,j}$. For each two fields $F(p,\bar z)\in{\cal
H}_{m,j}$, $G(p,\bar z)\in{\cal H}_{m,j}$ it reads
\renewcommand{\theequation}{\arabic{section}.\arabic{equation}a}
\begin{equation}
\langle F|G\rangle_{m,j}=m^{2\kappa+2j-1}{\cal
N}(\kappa,j) \int\frac{d{\bf p}}{p^0}\int\limits_{|z|<1}
\frac{dzd\bar z}{2\pi i} \frac{(1-z\bar z)\lefteqn{{}^{-2-2j}}}{|(p,n)|
\lefteqn{{}^{2\kappa+2j}}}
\qquad\overline{F(p,\bar z)}G(p,\bar z)\,,\label{4.30}
\end{equation}
\renewcommand{\theequation}{\arabic{section}.\arabic{equation}}%
where $\kappa>1/2$ is a real parameter and
\begin{displaymath}
\displaystyle
[{\cal N}(\kappa,j)]^{-1}=\int\limits_{|z|<1}
\frac{dzd\bar z}{2\pi i} \frac{(1-z\bar z)\lefteqn{{}^{-2-2j}}}{(1+z\bar z)
\lefteqn{{}^{2\kappa+2j}}}\qquad=\frac{1}{2^{2j+1}}\sum\limits_{k=0}^{2j}
\frac{(2j)!}{k!(2j-k)!(k+2\kappa-1)}
\end{displaymath}
is a normalization constant.
The operators (\ref{4.27}) are Hermitian with respect to the inner product.
Furthermore, parameter $\kappa$ is, in fact, inessential, because
the UIR's, corresponding different $\kappa$ in Eq.~(\ref{4.30}) and
the same $m,j$, are unitary equivalent to each other. It is explicitly
seen, if one takes the integral (\ref{4.30}) over ${\cal L}$ with account of
the expansion (\ref{4.22}) for the wave functions.
Then the inner product~(\ref{4.30}) transforms to the following form
\addtocounter{equation}{-1}
\renewcommand{\theequation}{\arabic{section}.\arabic{equation}b}
\begin{equation}
\langle F_j|G_j\rangle_{m,j}=\frac{1}{m}\int\frac{d{\bf p}}{p^0}\overline{
F^{\,\alpha_1\alpha_2\dots\alpha_{2j}}(p)}
G_{\,\alpha_1\alpha_2\dots\alpha_{2j}}(p)
\,,\label{4.31}
\end{equation}
\renewcommand{\theequation}{\arabic{section}.\arabic{equation}}%
\sbox{\innprod}{(\theequation)}\addtocounter{equation}{1}%
which is independent from $\kappa$.

The superextension appears immediately. We need only to consider the
space ${\cal H}_{m,s,b}$ introduced in subsec.~2 and to put $s+1=-j\leq0$
being a (half)integer number. The wave function has the following component
expansion
\begin{eqnarray}
F(p,\bar{\mit\Gamma})&=&
F_{0\,\alpha_1\alpha_2\dots\alpha_{2j+2}}(p)\bar z^{\alpha_1}
\bar z^{\alpha_2}\dots\bar z^{\alpha_{2j+2}}\nonumber\\&&+
i\sqrt{(j+1)(1+b)}\bar\theta F_{1\,\alpha_1\alpha_2\dots\alpha_{2j+1}}(p)\bar
z^{\alpha_1} \bar z^{\alpha_2}\dots\bar z^{\alpha_{2j+1}}\label{4.32}\\&&+
i\sqrt{(j+1)(1-b)}\bar\chi F_{2\,\alpha_1\alpha_2\dots\alpha_{2j+1}}(p)\bar
z^{\alpha_1} \bar z^{\alpha_2}\dots\bar z^{\alpha_{2j+1}}\nonumber\\&&
+\sqrt{(j+1/2)(j+1)(1-b^2)}\bar\theta\bar\chi F_{3\,\alpha_1\alpha_2\dots
\alpha_{2j}}(p)\bar z^{\alpha_1}\bar z^{\alpha_2}\dots\bar z^{\alpha_{2j}}
\nonumber
\end{eqnarray}
and it is subjected the wave equations (\ref{4.16}). The latter reduce to
the ``Dirac equations'' (\ref{4.29}) for each of the component
$F_{\imath\,\alpha_1 \alpha_2\dots\alpha_{j_i}}\,,i=0,1,2,3$. The
Hermitian inner product in ${\cal H}_{m,j,b}\equiv{\cal H}_{m,s,b}$ may be
introduced by analogy with Eq.~(\ref{4.2b}) and for each two
$F(p,\bar{\mit\Gamma})\,, G(p,\bar{\mit\Gamma})\in {\cal H}_{m,j,b}$ is
expressed in terms of the inner product~\usebox{\innprod}
\begin{equation}
(F|G)=\langle F_0|G_0\rangle_{m,j+1}+\langle F_1|G_1\rangle_{m,j+1/2}+
\langle F_2|G_2\rangle_{m,j+1/2}+\langle F_3|G_3\rangle_{m,j}\,.\label{4.33}
\end{equation}
It is a matter of  direct verification to prove that the operators
(\ref{4.15}), which generate the representation of the N=2 Poincar\'e
superalgebra, are truly Hermitian with respect to this inner product.
Moreover, the BPS bound $|b|\leq1$ and the reality of the renormalized
value of ${\rm q}$ (\ref{4.19}) by $s\leq-1$ provides the necessary and
sufficient conditions for operators (\ref{4.15}) to be Hermitian.

Thus, the quantization of the model reproduces the Hilbert space of
the (half)inte\-ger superspin superparticle. Each component of the wave
function (\ref{4.32}) describes a particle with fixed spin. N=2 supersymmetry
unifies four particles of the equal mass $m$ and of the (half)integer spins
$j+1,j+1/2,j+1/2,j,(j\geq0)$ in the irreducible superquartet.

%%%%%%%%%%%%%%%%%%%%%%%%%%%%%%%%%%%%%%%%%%%%%%%%%%%%%%%%%%%%%%%%%%%%%%%%%
\subsection[N=2 hypermultiplet]{N=2 hypermultiplet} \label{sbs4.5} %%%%%%
\hspace*{\parindent}\nopagebreak %%%%%%%%%%%%%%%%%%%%%%%%%%%%%%%%%%%%%%%%
%%%%%%%%%%%%%%%%%%%%%%%%%%%%%%%%%%%%%%%%%%%%%%%%%%%%%%%%%%%%%%%%%%%%%%%%%
The theory is strongly simplified in the BPS limit when $|b|=1$. The
classical phase superspace appears to be ${\cal M}^{8|2}=T^\ast({\rm
R}^{1,2})\times{\cal L}^{1|1}$, where the atypical $\rm OSp(2|2)$
co-orbit~${\cal L}^{1|1}$ of complex dimension $1/1$ plays the role of the
inner supermanifold associated to the superparticle superspin. The
atypical coadjoint orbit of the $\rm OSp(2|2)$ substitutes the typical one
in the BPS limit.

The supermanifold ${\cal M}^{8|2}$ serves the extended phase superspace of
the N=1, D=3 superparticle allowing the hidden N=2 supersymmetry.
Therefore, the quantization of the N=2 superparticle reduces in the BPS
limit to the one of the N=1 superparticle. Following the same method we
may combine the canonical Dirac procedure and the geometric quantization.
The respective theory of N=1 superanyon has been considered
earlier~\cite{GKL2} and it results in the description of N=1, D=3
supersymmetric doublet of anyons in terms of the fields carrying the
atypical UIR's of the $\rm OSp(2|2)$.  Moreover, it is shown in
Ref.~\cite{GKL2} that the N=1 superdoublet allows extended N=2 SUSY and it
can be treated as the N=2 hypermultiplet of anyons. It is
exactly N=2 fractional spin superparticle emerged in the BPS limit of the
N=2 model suggested in the present paper.  Let us briefly consider some
details of this limiting case.

One can see, that the consideration of this Section is easily modified
to the case, which saturates the BPS bound, when $|b|=1$. Consider, for
instance, $b=1$.  Then the wave function~(\ref{4.14}) does not depend on
$\chi$. It is equivalent to the vanishing of half of the odd variables in
the BPS limit mentioned at the classical level. The generators~(\ref{4.15})
of the N=2 Poincar\'e superalgebra reduce to \begin{equation}
\begin{array}{l}\displaystyle
{\widehat{\cal J}}_a=-i\epsilon_{abc}p^b\frac\partial{\partial p_c}
+{\widehat J}{}_a\qquad\qquad {\widehat{\cal P}}_a=p_a\qquad\qquad
{\widehat{\cal Z}}=m\\
\displaystyle
{\widehat{\cal Q}}{}^1_\alpha=(ip_{\alpha\beta}{\widehat
W}{}^\beta+m{\widehat{\tilde W}}{}_\alpha) \qquad
{\widehat{\cal Q}}{}^2_\alpha=(ip_{\alpha\beta}{\widehat{\tilde W}}{}^\beta
-m {\widehat W}{}_\alpha)\,,
\end{array} \label{4.34}
\end{equation}
where
\begin{eqnarray}&\displaystyle
{\widehat J}{}_a=-\bar\xi_a\bar\partial-(\bar\partial\bar\xi_a)
\left(1+\frac12\bar\theta
\frac{\partial}{\partial\bar\theta}\right)&
\hspace*{-2cm}\label{4.35}\\&\displaystyle\hspace*{-6mm}
\sqrt{ms}{\widehat W}^\alpha=\bar\theta(\bar
z^\alpha\bar\partial+2s(\bar\partial \bar z^\alpha))-\bar
z^\alpha\frac\partial{\partial\bar\theta}\qquad \sqrt{ms}{\widehat{\tilde
W}}{}^\alpha=-i\bar\theta(\bar z^\alpha\bar\partial+2s(\bar\partial \bar
z^\alpha))-i\bar
z^\alpha\lefteqn{\frac\partial{\partial\bar\theta}\,.}&\nonumber
\end{eqnarray}

The operators (\ref{4.35}) together with one more scalar $\rm U(1)$-generator
\begin{displaymath}
{\widehat B}\equiv{\widehat P_4}{}-s=\frac12\bar\theta\frac\partial{\partial
\bar\theta}-s
\end{displaymath}
form an irreducible representation of the $\rm OSp(2|2)$, which is unitary
for $s>0$ (it is an atypical UIR of the $\rm OSp(2|2)$ mentioned above)
and it is finite dimensional for~$s=-j$, $j$ being positive (half)integer.
The expressions (\ref{4.35}) can also be obtained by straightforward
geometric quantization on~${\cal L}^{1|1}$~\cite{Balant,Grad}.

The operators (\ref{4.34}) are {\it linear\/} in the generators of the inner
$\rm osp(2|2)$ superalgebra. Therefore, they generate N=2 Poincar\'e
superalgebra with central charge ${\cal Z}=m$ immediately,
without any corrections in $1/s$, and the renormalization is not required
for the case. The wave equations (\ref{4.16}) hold their form in the BPS
limit and the inner products are given by Eqs.~(\ref{4.2b}), (\ref{4.20}),
(\ref{4.33}), where one should account for the vanishing of the last two
components of the wave functions in the expansion~(\ref{4.14}).

The peculiarities of the model  mentioned, mean that we obtain in the BPS
limit an adequate description of the N=2 particle hypermultiplet. In
particular, we have a natural smooth reduction for  both the Poincar\'e
supersymmetry and the internal one. The comparison of the classical
mechanics given in subsec.~III.6 and of the presented quantum theory
demonstrates the direct relationship between the contraction of the
classical phase superspace $T^\ast({\rm R}^{1,2})\times{\cal L}^{1|2}$ to
$T^\ast({\rm R}^{1,2})\times{\cal L}^{1|1}$ and the shortening of
the~N=2 particle supermultiplet to the N=2 hypermultiplet in the BPS limit.

%%%%%%%%%%%%%%%%%%%%%%%%%%%%%%%%%%%%%%%%%%%%%%%%%%%%%%%%%%%%%%%%%%%%%%%%%%%
\section[\hspace{4mm}Summary and outlook]
{\normalsize\bf\hspace{-5mm}SUMMARY AND DISCUSSION}\label{s5}
\hspace*{\parindent}%
%%%%%%%%%%%%%%%%%%%%%%%%%%%%%%%%%%%%%%%%%%%%%%%%%%%%%%%%%%%%%%%%%%%%%%%%%%%
In the present paper we have constructed the consistent first quantized
theory of N=2, D=3 superanyon as well as the one of massive superparticle of
the habitual (half)integer superspin. The starting point for the quantization
is the classical model of the superparticle in the nonlinear phase superspace
${\cal M}^{8|4}=T^\ast({\rm R}^{1,2})\times{\cal L}^{1|2}$, that is
different from the standard approach.

A traditional viewpoint in the construction of the spinning particle
models~\cite{Frydr} is to describe the spinning degrees of freedom by
some variables being simultaneously translation invariant and
Lorentz covariant (as usual, those are Lorentz vectors or spinors). Such
variables parametrize some linear space $L$ and then the extended phase
space is chosen to be ${\cal M}=T^\ast({\rm R}^{1,D-1})\times L$ or
${\cal M}=T^\ast({\rm R}^{1,D-1}\times L)$. The only difference for
superparticles is to replace D-dimensional Minkowski space ${\rm R}^{1,D-1}$
by the respective superspace. The advantage of the covariant
(super)space ${\cal M}$ is in the linear (``covariant'') action of the
Poincar\'e supergroup. In this approach, however, an embedding of the
(super)particle physical space ${\cal O}$ (that is the underlying coadjoint
orbit) in the covariant phase (super)space may be ambiguous.
Moreover, it is a common usage in this approach to give little attention
to the geometry underlying the embedding ${\cal O}\to {\cal M}$.

We have demonstrated that the nonlinear phase superspace
${\cal M}^{8|4}=T^\ast({\rm R}^{1,2})\times{\cal L}^{1|2}$ of~D=3
spinning superparticle has the following remarkable features:

(i) The embedding of an appropriate coadjoint ${\cal O}_{m,s,b}$ orbit,
being associated to the N=2, D=3 superparticle of arbitrary fixed mass $m>0$,
superspin $s\neq0$ and central charge $mb$ ( $|b|<1$ in ${\cal M}^{8|4}$), \,
is realized by two constraints, which provide the identical conservation
of any Casimir function of the Hamiltonian Poincar\'e superalgebra. These
constraints have transparent geometric origin and, after quantization, they
are converted into wave equations of the superparticle in a natural way.

(ii) The `inner' subsupermanifold ${\cal L}^{1|2}$ of ${\cal M}^{8|4}$
appears to be in itself the coadjoint orbit for some supergroups.
${\cal L}^{1|2}$ is shown to be symplectomorphic to the
K\"ahler homogeneous superspace of the supergroup $\rm SU(1,1|2)$ or its
subsupergroup $\rm OSp(2|2)$. In this sense the model admits the second
supersymmetry ($\rm SU(1,1|2)$ SUSY) along with the original Poincar\'e one.

To describe the superparticle in a standard way, it is convenient, starting
from an ordinary particle living in ${\rm R}^{1,D-1}$, to extend the
geometry of the Minkowski space to the supergeometry of the respective
Minkowski superspace. We have found an alternative way,
at least for dimension D=3. The intrinsic structure of D=3 spinning particle
may be described in terms of the Lobachevsky geometry. To introduce the
supersymmetry we may extend the inner manifold, going to the
Lobachevsky supergeometry.

The following interpretation is admissible. D=3 particle lives in an
ordinary Minkowski space ${\rm R}^{1,2}$. In addition the {\it super\/}spin
degrees of freedom are associated to its internal structure and generate the
internal phase superspace ${\cal L}^{1|2}$ with an inherent~${\rm
SU(1,1|2)}$ supersymmetry, which is different from the Poincar\'e
(super)symmetry.

(iii) The BPS limit of the model looks slightly different if compared to
the standard picture for the superparticle~\cite{AzcLuk}. In the standard
approach the extended superparticle model with central charge reveals
the generation of new gauge degrees of freedom in the BPS limit and it
corresponds to an appropriate reduction for the physical phase space.
Furthermore, it is usually impossible to impose the gauge in a covariant way
and then to forget about the nonphysical variables. In contrast to the
standard approach, the phase superspace $T^\ast({\rm R}^{1,2})\times{\cal
L}^{1|2}$ can be explicitly truncated to $T^\ast({\rm R}^{1,2})\times{\cal
L}^{1|1}$ in the BPS limit. In this case the inner N=2 Lobachevsky
supergeometry reduces to the N=1 one, while the gauge variables drop out
from the theory at all.  The reduction of the phase superspace in the
classical mechanics is directly related to the shortening of the particle
supermultiplet in quantum theory considered in the BPS limit.

(iv) We suggest nontrivial quantization of the superparticle in
the extended phase superspace ${\cal M}^{8|4}$, which combines the
canonical quantization in $T^\ast({\rm R}^{1,2})$ and the Berezin
quantization in the inner phase superspace ${\cal L}^{1|2}$. This
quantization scheme leads naturally to the fields carrying infinite
dimensional or finite dimensional representation of the supergroup $\rm
SU(1,1|2)$ depending on the fractional or habitual
(half)integer value of spin. The results are completely consistent with the
previous known description of D=3 nonsupersymmetric particles as mechanical
systems~\cite{JackNair,CorPly}.

Surprisingly, there are two, unitary equivalent to one another, series of N=2
supercharges in quantum theory, which correspond to different possibilities
for the parameter $\rm q$ in Eq.~(\ref{4.19}). Only one of them, namely $\rm
q_-$, is related directly to a conventional classical limit. Another
possible value $\rm q_+$ is shown to be related to the special properties
(\ref{paritycomm}) of the parity operator (\ref{parityop}). However, the
classical counterpart of this parity structure remains unclear, and the
origin of the second possible value for $\rm q$ may seem enigmatic. Notice
that the parity operator generates the structure of the deformed
Heisenberg algebra in Hilbert space of anyon~\cite{Ply1} or N=1
superanyon~\cite{GKL2}. It would be interesting to understand, what
is a geometry behind the parity operator for N=2 superanyon.

The significance of this one-particle theory may vary, in particular,
depending
on the possibility of an efficient second quantization of the model.
One of the problems here is to construct a Lagrangian of the theory,
which leads to the one-particle wave equation we have deduced from
the classical mechanical action. The first
step of construction may be to present two independent wave equations of
superanyon (like Eqs.~(\ref{4.16})) in the form of one spinor equation, when
the mass  and spin shell fixing conditions may emerge as integrability
conditions. It is known that the similar construction  for anyons gives a
simple action functional~\cite{SorVol,Ply1}, which may be relevant for the
second quantization of fractional spin particles. An adequate
superextension (at least for N=1) may be constructed probably using the
representations of the $\rm su(1,1|2)$ superalgebra in the same way, as the
spinor set of the anyon wave equations was constructed in Ref.~\cite{Ply1}
using the atypical UIR's of~$\rm osp(2|2)$. In this
connection it should be noted that the exploitation of the atypical UIR's
of the~$\rm osp(2|2)$ superalgebra and of the deformed Heisenberg algebra
produces the linear set of spinor wave equations of N=1, D=3 superparticle
only for special (half)integer~$j=1/2$ and~$j=1$ values of the
superspin~\cite{Ply1}.

And, of course, the consistent interaction of (super)anyons remains an
intriguing problem. Even in the first quantized theory the
suggested approach to the description of anyon, being attempted for the
extension to an interaction with an external field, implies (in
the framework of minimal phase space) a perturbative representation
for nonlinear commutation relations in terms of a series in powers of the
field strengths~\cite{ChouNairPol}. In particular, it is unclear whether any
 consistent generalization exists for the wave equations of (super)anyons
obtained in this paper in the presence of arbitrary external fields.

%%%%%%%%%%%%%%%%%%%%%%%%%%%%%%%%%%%%%%%%%%%%%%%%%%%%%%%%%%%%%%%%%%%%%%%%%%%
\section*{\normalsize\bf ACKNOWLEDGMENTS}\label{s6}\hspace*{\parindent}
S.L.L.~is thankful to Professor S.~Randjbar-Daemi for the hospitality
at the High Energy Section, ICTP, where the manuscript of the paper has
been completed.

This work is partially supported by Grants of the International Soros Science
Education Program a98-166, d-98-932 and by the RBRF Grant  98-02-16261. S.L.L.
was supported in part by the Joint INTAS-RBRF Grant 95-829.
I.V.G. is supported partially by the INTAS Grant 96-0457
within the research program of the International Center for Fundamental
Physics in Moscow.

%%%%%%%%%%%%%%%%%%%%%%%%%%%%%%%%%%%%%%%%%%%%%%%%%%%%%%%%%%%%%%%%%%%%%%%%%%%

%%%%%%%%%%%%%%%%%%%%%%%%%%%%%%%%%%%%%%%%%%%%%%%%%%%%%%%%%%%%%%%%%%%%%%%%%%
\appendix
\addcontentsline{toc}{section}{Appendix. {\rm Anticommutation relations and
renormalization}\protect\\ \hspace*{16mm}
{\rm for the N=2 Poincar\'e supercharges}}
\subsection*{Appendix. Anticommutation relations and
renormalization\\ \hspace*{27mm} for the N=2 Poincar\'e supercharges}%
\setcounter{equation}{0}\renewcommand{\theequation}{A.\arabic{equation}}
\hspace*{\parindent}%
%%%%%%%%%%%%%%%%%%%%%%%%%%%%%%%%%%%%%%%%%%%%%%%%%%%%%%%%%%%%%%%%%%%%%%%%%%
We give here the calculation of the anticommutator (\ref{4.18}) of N=2
Poincar\'e supercharges in more detail. The object is to show that the
closure of the Poincar\'e superalgebra requires a renormalization of the
parameter ${\rm q}$ entering in the definition of the supercharge's
(\ref{4.15}) and the renormalized value is one of given by
Eq.~(\ref{4.19}).

Before coming  into explicit formulas, it is helpful to introduce a
convenient notation, which is slightly different from that used in the paper.
First, we redefine the coefficients in the expansion (\ref{4.14}) and write
down the wave function in the form
\begin{equation}
F(p,\bar{\mit\Gamma})=F_0(p,\bar{z})+\bar\theta F_1(p,\bar{z})
+\bar\chi F_2(p,\bar{z})+\bar\theta\bar\chi F_3(p,\bar{z})\,.\label{A1}
\end{equation}
Hereafter, we shall represent the wave function $F\in{\cal H}$ as a
four-column
\begin{equation}
F=\left(\begin{array}{c}F_0\\F_1\\F_2\\F_3\end{array}\right)\,,\label{A2}
\end{equation}
where the components $F_i\,, i=0,1,2,3$ depend on $p$ and $\bar z$. In
these terms each linear operator $\widehat A$ in ${\cal H}$ will
be presented by
a matrix $({\widehat A}{}_i^j)$ of dimension $4\times4$, whose elements are
operators acting on the components. The matrix elements
$({\widehat A}{}_i^j)$ are defined by
the rule $({\widehat A} F)_i=\sum_{j=0}^3{\widehat A}{}_i^jF_j$. The matrix
representation gives a convenient tool for the explicit calculations performed
below.

Second, we take $q^\prime=1-{\rm q}$ to squeeze the notations.

Let us note that the Poincar\'e supercharge's (\ref{4.15}) may be identically
presented in the following form
\begin{equation}
{\widehat{\cal Q}}{}^{1\,\alpha}=ip^\alpha{}_\beta{\widehat W}^\beta_q+
m{\widehat{\tilde W}}{}^\alpha_q \, ,\qquad
{\widehat{\cal Q}}{}^{2\,\alpha}=ip^\alpha{}_\beta{\widehat V}^\beta_q+
m{\widehat{\tilde V}}{}^\alpha_q\,, \label{A3}
\end{equation}
where
\begin{eqnarray}
\displaystyle
{\widehat W}^\alpha_q=\frac{1}{2\sqrt{ms}}\left(
\begin{array}{rccc}
\displaystyle 0\qquad&\displaystyle-\bar z^\alpha&\displaystyle-i\bar
z^\alpha&0\\ \displaystyle\frac{1+b}2{\hat Q}^\alpha_0&\displaystyle 0&0
&iq^\prime\bar z^\alpha\\
\displaystyle -i\frac{1-b}2{\hat Q}^\alpha_0&\displaystyle 0&0
&-q^\prime\bar z^\alpha\\
0\qquad&\displaystyle iq^\prime\frac{1-b}2{\hat Q}^\alpha_1&\displaystyle
q^\prime\frac{1+b}2{\hat Q}^\alpha_1&0
\end{array}\right)&&\nonumber\\[5mm] \displaystyle
{\widehat{\tilde W}}{}^\alpha_q=\frac{1}{2\sqrt{ms}}\left(
\begin{array}{rccc}
\displaystyle 0\qquad&\displaystyle-i\bar z^\alpha&\displaystyle\bar
z^\alpha&0\\ \displaystyle-i\frac{1+b}2{\hat Q}^\alpha_0&\displaystyle 0&0
&-q^\prime\bar z^\alpha\\
\displaystyle-\frac{1-b}2{\hat Q}^\alpha_0&\displaystyle 0&0
&-iq^\prime\bar z^\alpha\\
0\qquad&\displaystyle q^\prime\frac{1-b}2{\hat Q}^\alpha_1&\displaystyle
-iq^\prime\frac{1+b}2{\hat Q}^\alpha_1&0
\end{array}\right)&&\nonumber\\&&\label{A4}\\
\displaystyle
{\widehat V}^\alpha_q=\frac{1}{2\sqrt{ms}}\left(
\begin{array}{rccc}
\displaystyle0\qquad&\displaystyle-i\bar z^\alpha&\displaystyle-\bar
z^\alpha&0\\ \displaystyle-i\frac{1+b}2{\hat Q}^\alpha_0&\displaystyle 0&0
&q^\prime\bar z^\alpha\\
\displaystyle\frac{1-b}2{\hat Q}^\alpha_0&\displaystyle 0&0
&-iq^\prime\bar z^\alpha\\
0\qquad&\displaystyle -q^\prime\frac{1-b}2{\hat Q}^\alpha_1&\displaystyle
-iq^\prime\frac{1+b}2{\hat Q}^\alpha_1&0
\end{array}\right)&&\nonumber\\[5mm] \displaystyle
{\widehat{\tilde V}}{}^\alpha_q=\frac{1}{2\sqrt{ms}}\left(
\begin{array}{rccc}\displaystyle
0\qquad&\displaystyle\bar z^\alpha&\displaystyle-i\bar z^\alpha&0\\
\displaystyle-\frac{1+b}2{\hat Q}^\alpha_0&\displaystyle 0&0
&iq^\prime\bar z^\alpha\\
\displaystyle-i\frac{1-b}2{\hat Q}^\alpha_0&\displaystyle 0&0
&q^\prime\bar z^\alpha\\
0\qquad&\displaystyle iq^\prime\frac{1-b}2{\hat Q}^\alpha_1&\displaystyle
-q^\prime\frac{1+b}2{\hat Q}^\alpha_1&0
\end{array}\right)\,,&&\nonumber
\end{eqnarray}
and ${\hat Q}^\alpha_\epsilon=\bar z^\alpha\bar\partial+(2s+\epsilon)
(\bar\partial\bar z^\alpha)\,;\epsilon=0,1$. When $q^\prime=1$ we have
${\widehat W}^\alpha_0\equiv{\widehat W}^\alpha
\,,{\widehat{\tilde W}}{}^\alpha_0\equiv{\widehat{\tilde W}}{}^\alpha$ and
similar identities for $\widehat V$'s.

The calculations of the anticommutators of N=2 supercharges give
\begin{displaymath}
[{\widehat{\cal Q}}{}^{1\,\alpha}\,,{\widehat{\cal
Q}}^{1\,\beta}]_+=2p^{\alpha\beta}- \frac{1}{4ms}{\widehat X}_+{\widehat
J}^{\alpha\beta}(p^2+m^2)+ \frac{1}{2ms}{\widehat X}_+p^{\alpha\beta}((p,
\widehat J)-m\widehat P_4-ms)+ p^{\alpha\beta}{\widehat O}_+
\end{displaymath}
\begin{displaymath}
[{\widehat{\cal Q}}{}^{2\,\alpha}\,,{\widehat{\cal
Q}}{}^{2\,\beta}]_+=2p^{\alpha\beta}- \frac{1}{4ms}{\widehat X}_-{\widehat
J}^{\alpha\beta}(p^2+m^2)+ \frac{1}{2ms}{\widehat X}_-p^{\alpha\beta}((p,
\widehat J)-m\widehat P_4-ms)+ p^{\alpha\beta}{\widehat O}_- \,,
\end{displaymath}
\begin{eqnarray}\displaystyle
[{\widehat{\cal Q}}{}^{2\,\alpha}\,,{\widehat{\cal Q}}{}^{1\,\beta}]_+&
=&\displaystyle-2imb
\epsilon^{\alpha\beta}-\frac{i}{8ms}(\widehat X_1\epsilon^{\alpha\beta}
+i\widehat X_0\widehat J^{\alpha\beta})(p^2+m^2)\label{A5}\\&&\displaystyle
-\frac{i}{4ms}(m\widehat X_2\epsilon^{\alpha\beta}
+i\widehat X_0p^{\alpha\beta})
((p,\widehat J)-m\widehat P_4-ms)+m\widehat O^{\alpha\beta}\nonumber\,,
\end{eqnarray}
where ${\widehat J}^{\alpha\beta}=(\gamma^a)^{\alpha\beta}{\widehat J}_a$\,,
\begin{displaymath}
\widehat X_\pm=\left(
\begin{array}{cccc}
2&0&0&0\\
0&(1+b)+q^{\prime\,2}(1-b)&\pm i(1+b)(1-q^{\prime\,2})&0\\
0&\mp i(1-b)(1-q^{\prime\,2})&(1-b)+q^{\prime\,2}(1+b)&0\\
0&0&0&2q^{\prime2}
\end{array}\right)
\end{displaymath}
\begin{displaymath}
\widehat X_0=2(1-q^{\prime\,2})\left(
\begin{array}{cccc}
0&0&0&0\\
0&0&1+b&0\\
0&1-b&0&0\\
0&0&0&0
\end{array}\right)\,,
\end{displaymath}
$\widehat X_{1,2}$ are diagonal operators
\begin{eqnarray}
\displaystyle
\widehat X_1F&=&\displaystyle 4bsF_0+[(1+b)(2s-1)-q^{\prime\,2}(1-b)(2s+1)]F_1
\nonumber\\&&
\displaystyle-[(1-b)(2s-1)-q^{\prime\,2}(1+b)(2s+1)]F_2+4bsq^{\prime\,2}F_3
\nonumber\\
\displaystyle
\widehat X_2F&=&\displaystyle 4bF_0+2[(1+b)-q^{\prime\,2}(1-b)]F_1
-2[(1-b)-q^{\prime\,2}(1+b)]F_2+4bq^{\prime\,2}F_3\,, \nonumber
\end{eqnarray}
and
\begin{equation}
\begin{array}{c}\displaystyle
\widehat O_\pm=\left(q^{\prime\,2}\frac{2s+1}{2s}-1\right)\left(
\begin{array}{cccc}
0&0&0&0\\
0&1-b&\mp i(1+b)&0\\
0&\pm i(1-b)&1+b&0\\
0&0&0&2
\end{array}\right)\\[10mm] \displaystyle
\widehat O^{\alpha\beta}=\left(q^{\prime\,2}\frac{2s+1}{2s}-1\right)\left(
\begin{array}{cccc}
0&0&0&0\\
0&i(1-b)\epsilon^{\alpha\beta}&\displaystyle -\frac{(1+b)}{m}p^{\alpha\beta}
&0\\
0&\displaystyle -\frac{(1-b)}{m}p^{\alpha\beta}&-i(1+b)\epsilon^{\alpha\beta}
&0\\0&0&0&-2ib\epsilon^{\alpha\beta}
\end{array}\right)\,.
\end{array}\label{A6}
\end{equation}

What we have obtained in Eqs.~(\ref{A5}) is in fact a detailed notation
for Eqs.~(4.15). In this notation the structure of the anticommutation
relations becomes transparent and a number of helpful features stands out.
Let us consider the physical subspace in~${\cal H}$, which is generated by
solutions of the wave equations~(\ref{4.16}). Then the relations~(\ref{A5})
simplify themselves and we have
\begin{equation}
[\widehat{\cal Q}^I_\alpha,\widehat{\cal Q}^J_\beta]_+=
2\delta^{IJ}p_{\alpha\beta}-2imb\epsilon^{IJ}\epsilon_{\alpha\beta}
-m\delta^{IJ}\epsilon_{\alpha\beta}\widehat O_\pm-m\epsilon^{IJ}\widehat
O_{\alpha\beta}\,.\label{A7}
\end{equation}
The presence of the operators $\widehat O_\pm$ and $\widehat
O_{\alpha\beta}$ breaks off N=2 Poincar\'e superalgebra in general.
In the neighborhood of the value $q^\prime=1-1/4s$, being derived
from the combination of the classical mechanics and of the correspondence
rules, the operators~$\widehat O_\pm,\widehat O_{\alpha\beta}$ should be
treated as the corrections of order $s^{-2}$. However, it contains more.
The explicit expressions (\ref{A6}) show immediately that in the case of
$s\neq-1/2$ the renormalization of $q^\prime$ is possible, which provides
the identical vanishing of any corrections and the closure of the Poincar\'e
superalgebra on shell.  The renormalized values are
presented~by~Eq.~(\ref{4.19}).

The following observation is that the anticommutators obtained are invariant
under the substitution $q^\prime\to-q^\prime$ (that is ${\rm q}\to2-q$). As
mentioned in Subsec.~4.2, this invariance comes from the degenerate N=4
supersymmetry and from specific properties of the parity operator
(\ref{parityop}).

Finally, the structure of the anticommutators (\ref{A5}) changes drastically
in the BPS limit $|b|=1$. Consider, for instance, $b=1$. In this case, the two
latter components  of the wave function (\ref{A1}) vanish, $F_2=F_3\equiv0$
(see Eq.\ (\ref{4.14})). However, when $b=1$, in the linear subspace
$F_2=F_3\equiv0$ the parameter $q$ becomes inessential and the corrections
(\ref{A6}) vanish identically. Thus, we do not need any renormalization in
the BPS limit, which corresponds to the N=1 superparticle, which was considered
earlier in Ref.~\cite{GKL2}.


\begin{thebibliography}{MM}
\addcontentsline{toc}{section}{\refname}
\bibitem{Laugh}
R.B. Laughlin, Phys. Rev. Lett. {\bf 50}, 1395 (1983).
\bibitem{Wilczekbook}
F. Wilczek, in {\it Fractional Statistics and Anyon Superconductivity,\/}
(World Scientific, Singapore, 1990), edited by F. Wilczek.
\bibitem{LeiMyr}
J. M. Leinaas and J. Myrheim, Nuovo Cim. {\bf B37}, 1 (1977).
\bibitem{Wilc}
F. Wilczek, Phys. Rev. Lett. {\bf 48}, 1144 (1982);
{\it ibid.} {\bf 49}, 957 (1982);
\bibitem{Goldin}
G. A. Goldin, R. Menikoff and D. H. Sharp, J. Math. Phys. {\bf 21}, 650
(1980); {\it ibid.} {\bf 22}, 1664 (1981).
\bibitem{JackNair}
R. Jackiw and V. P. Nair, Phys. Rev. {\bf D43}, 1933 (1991).
\bibitem{Forte}
S. Forte and T. Jolicoeur, Nucl. Phys. {\bf B350}, 589 (1991).
\bibitem{SorVol}
D.P. Sorokin and D.V. Volkov, Nucl. Phys. {\bf B409}, 547 (1993).
\bibitem{Chern-Sim}
G.W. Semenoff, Phys. Rev. Lett. {\bf 61}, 517 (1988);\\
J. Fr\"{o}hlich and P.A. Marchetti, Lett. Math. Phys. {\bf 16}, 347 (1988);
Commun. Math. Phys. {\bf 121}, 177 (1988);
Nucl. Phys. {\bf B356}, 533 (1991), and references therein.
\bibitem{HlousSpect}
Z. Hlousek and D. Spector, Nucl. Phys. {\bf B344}, 763 (1990).
\bibitem{Balach}
A. Balachandran, G. Marmo, B. Skagerstam and A. Stern, {\it Gauge Symmetries
and Fibre Bundles,\/} (Springer, Berlin, 1983).
\bibitem{Plyush}
M.S. Plyushchay, Phys. Lett. {\bf B248}, 107 (1990); Int. J. Mod. Phys.
{\bf A7}, 7045 (1992).
\bibitem{CorPly}
J. L. Cort\'es and M. S. Plyushchay, Mod. Phys. Lett. {\bf A10}, 409 (1995);
Int. J. Mod. Phys. {\bf A11}, 3331 (1996).
\bibitem{Ghosh95}
S. Ghosh, Phys. Rev. {\bf D51}, 5827 (1995).
\bibitem{GKL1}
I. V. Gorbunov, S. M. Kuzenko and S. L. Lyakhovich, Int. J. Mod. Phys.
{\bf A12}, 4199 (1997).
\bibitem{Ners}
A. Nersesian, Mod. Phys. Lett. {\bf A12}, 1783 (1997).
\bibitem{ChouNairPol}
C. Chou, V.P. Nair and A.P. Polychronakos, Phys. Lett. {\bf B304}, 105 (1993).
\bibitem{Chou}
C. Chou, Phys. Lett. {\bf B323}, 147 (1994).
\bibitem{GhoshMukh}
S. Ghosh and S. Mukhopadhyay, Phys. Rev. {\bf D51}, 6843 (1995).
\bibitem{BenGhosh}
N.Banerjee and S. Ghosh, Phys. Rev. {\bf D52}, 6130 (1995).
\bibitem{Ghosh97.1}
S. Ghosh, J. Phys. {\bf A30}, L821 (1997).
%{\it Screening in anyon gas,\/} preprint hep-th/9707109.
\bibitem{Ghosh97.2}
S. Ghosh, {\it Planar two-particle Coulomb interaction: classical and
quantum aspects,\/} preprint hep-th/9708045.
\bibitem{LyakSegShar}
S. L. Lyakhovich, A. Yu. Segal and A. A. Sharapov, Phys. Rev. {\bf D54}, 5223
(1996).
\bibitem{LShSh98} S. L. Lyakhovich, A. A. Sharapov and K. M. Shekhter.
{\it Massive spinning particle in any dimension. 1.~Integer spins,}
preprint hep-th/9805020, to appear Nucl. Phys. {\bf B}
\bibitem{GKL2}
I. V. Gorbunov, S. M. Kuzenko and S. L. Lyakhovich, Phys. Rev.
{\bf D56}, 3744 (1997).
\bibitem{Ply1}
M.S. Plyushchay, Ann. Phys. {\bf 245}, 339 (1996); Mod. Phys. Lett. {\bf
A11},
2953 (1996); Mod. Phys. Lett. {\bf A12}, 1153 (1997).
\bibitem{Sour}
J. M. Souriau, {\it Structure des Syst\'emes Dynamique,\/}
(Dunod, Paris, 1970).
\bibitem{Kost}
B. Kostant, {\it Lecture Notes in Mathematics,\/} {\bf 170}
(Springer-Verlag, New York, 1970), p.~87.
\bibitem{Kiril}
A. A. Kirillov, {\it Elements of the Representation Theory,\/}
(Mir, Moscow, 1974).
\bibitem{Berezin1}
F. A. Berezin, Math. USSR - Izvestija {\bf 38}, 1116, (1974); {\it ibid.\/}
{\bf 39}, 363 (1975).
\bibitem{Berezin2}
F. A. Berezin, Commun. Math. Phys. {\bf 40}, 153, (1975)
\bibitem{Wood}
N. M. J. Woodhouse, {\it Geometric quantization,\/} (Clarendon Press, Oxford,
1980).
\bibitem{MarnMart}
R. Marnelius and U. Martensson, Nucl. Phys., {\bf B335}, 395 (1991); Int. J.
Mod. Phys. {\bf A6}, 807 (1991).
\bibitem{Frydr}
A. Frydryszak, {\it Lagrangian models of particles with spin: the first
seventy years,\/} preprint hep-th/9601020,
and more than 80 references therein.
\bibitem{KuzLyakSeg}
S. M. Kuzenko, S. L. Lyakhovich and A. Yu. Segal, Int. J. Mod. Phys. {\bf
A10}, 1529 (1995); Phys. Lett. {\bf B348}, 421 (1995).
\bibitem{Balant}
A. B. Balantekin, H. A. Schmitt and P. Halse, J. Math. Phys. {\bf 30}, 274
(1989).
\bibitem{GL98} I.V.Gorbunov and  S.L.Lyakhovich, Phys. Lett. {\bf B422}, 293
(1998).
\bibitem{Grad}
El A.M. Gradechi and L. M. Nieto, Commun. Math. Phys. {\bf 175}, 521 (1996).
\bibitem{AzcLuk}
J. A. de Azc\'araga and J. Lukierski, Phys. Lett. {\bf B113}, 170 (1982);
Phys. Rev. {\bf D28}, 1337 (1983).
\bibitem{sohn}
P. Fayet, Nucl. Phys. {\bf B51}, 135 (1976);
M. F. Sohnius, Nucl. Phys. {\bf B138}, 109 (1978); Phys. Rep. {\bf 128}, 39
(1985).
\bibitem{Dictionary}
L. Frappat, A. Sciarrino and P. Sorba, {\it Dictionary on Lie
Superalgebras,\/} preprint ENSLAPP-AL-600/96, DSF-T-30/96, hep-th/9607161.
\bibitem{Brink}
L. Brink and M.B. Green, Phys. Lett. {\bf B106}, 393 (1981).
\bibitem{SpecHlou}
Z. Hlousek and D. Spector, Nucl. Phys. {\bf B370}, 143 (1992); {\it
ibid.\/} {\bf B397}, 173 (1993).
\bibitem{Perel}
M. A. Perelomov, {\it Generalized Coherent States,\/} (Nauka, Moscow, 1987);
\bibitem{BarMar}
D. Bar-Moshe and M.S. Marinov, in {\it Topics in statistical and
theoretical physics,\/} ed by R.L. Dobrushin et al.; J. Phys. {\bf A27}, 6287
(1994).
\bibitem{GelfGrVil}
I. M. Gelfand, M. I. Graiev and N. Ya. Vilenkin, {\em Generalized
Functions,\/} (Academic Press, New York, 1966) Vol.~5.

\end{thebibliography}
\end{document}